\documentstyle[12pt,epsf]{article}
\def\half{\frac{1}{2}}
\def\be{\begin{equation}}
\def\ba{\begin{eqnarray}}
\def\ea{\end{eqnarray}}   
\def\gr{\nabla}
\def\ee{\end{equation}}
\def\to{\rightarrow}
\def\tmu{$TH\epsilon\mu\;$}
\def\pd{\partial}
\def\nn{\not\!}
\def\sp{\overline\psi}
\def\api{\frac{\alpha}{\pi}}
\def\al{\alpha}
\def\uu{\vec u}
\def\pp{\vec p}

\def\nv{\hat n}

\renewcommand{\thesection}{\Roman{section}}

\begin{document}
\baselineskip=0.6cm
\begin{titlepage}
\begin{center}
{\Large\bf Testing the Equivalence Principle\vspace{0.4cm}\\
 by Lamb shift Energies}\vspace{1cm}\\
C. Alvarez
\footnotemark\footnotetext{E-mail:
calvarez@avatar.uwaterloo.ca}\\
and\\
R.B. Mann\footnotemark\footnotetext{E-mail:
mann@avatar.uwaterloo.ca}\\
Department of Physics \\
University of Waterloo\\
Waterloo, ONT N2L 3G1, Canada\\
\today\\
\end{center}

\begin{abstract}
The Einstein Equivalence Principle has as one of its implications
that the non-gravitational laws of physics are those of special
relativity in any local freely-falling frame. We consider
possible tests of this hypothesis for systems whose energies are due to
radiative corrections, {\it ie.} which arise
purely as a consequence of quantum field theoretic loop effects. 
Specifically, we
evaluate the Lamb shift transition (as given by the energy splitting 
between the
$2S_{1/2}$ and $2P_{1/2}$ atomic states) within the context of violations of
local position invariance and local Lorentz invariance,
as described by the $T H \epsilon\mu$ formalism.  We compute the associated
red shift and time dilation parameters, and discuss how
(high-precision) measurements 
of these quantities could provide  new 
information on the validity of the equivalence principle. 
\end{abstract}
\end{titlepage}
\section{Introduction}

The Einstein Equivalence Principle (EEP) is foundational to our
understanding of gravity. It states that (i) all test bodies fall with the
same acceleration regardless of their composition (the weak equivalence
principle, or WEP) and (ii) the outcome of any local nongravitational
test experiment is independent of the velocity 
and the spacetime orientation and location of the
(freely-falling) apparatus \cite{will1}.  Theories which obey the EEP, such as
general relativity and Brans-Dicke Theory are called metric
theories because they  endow spacetime with a metric $g_{\mu\nu}$ that
couples universally to all non-gravitational fields. 
Non-metric theories do not have this feature:  they break universality  by
coupling auxiliary gravitational fields directly to matter. In this
context a violation of the EEP means the breakdown of either
Local  Position Invariance (LPI) or Local Lorentz Invariance (LLI)
(or both) so that observers performing local
experiments could detect effects due to their position (if LPI is violated)
or their velocity (if LLI is violated) in an external
gravitational environment by using clocks and rods of differing
composition. Limits on LPI and LLI are set by gravitational red-shift  and
atomic physics experiments respectively \cite{redshift,PLC1,PLC}, each of which
compares relative frequencies of transitions between particular energy
levels that are sensitive to any potential LPI/LLI-violating effects.

The next generation of  gravitational experiments will  
significantly extend our 
current understanding of the empirical foundations of the EEP. 
A proposed E\"otv\"os experiment
in space, known as the Satellite Test of the Equivalence Principle (STEP)
attempts to test WEP to one part in $10^{17}$. The precision of gravitational 
red shift experiments could be
improved to one part in $10^{9}$  by placing a hydrogen maser 
clock on board Solar Probe, a proposed 
spacecraft (see ref. \cite{will1} and references therein).

The dominant form of energy governing the transitions these experiments probe
is nuclear electrostatic energy, although violations of WEP/EEP due to
other forms of energy (virtually all of which are associated with baryonic
matter) have also been estimated \cite{HW}.  However there exist many 
other physical systems, dominated by primarily non-baryonic
energies, for which the validity of the EEP is comparatively less  
well understood \cite{Hughes}.  Such systems include photons of differing
polarization \cite{Jody}, antimatter systems \cite{anti}, 
neutrinos \cite{utpal}, mesons \cite{Kenyon}, massive leptons, 
hypothesized dark matter, second and third generation matter, and
quantum vaccum energies. Indeed,
potential violations of the EEP due to vacuum energy shifts,
which are peculiarly quantum-mechanical in origin 
({\it i.e.} do not have a classical or semi-classical
description) provide an interesting empirical regime for gravitation
and quantum mechanics.  

In this paper we investigate the effects that EEP-violating couplings have 
on Lamb-shift transition energies.  Such transitions arise solely due to 
the radiative corrections inherent in quantum electrodynamics. A test of 
the EEP for this form of energy therefore provides
us with a qualitatively new empirical window of the foundations of 
gravitational theory.

The Lamb shift is the shift in energy levels of a Hydrogenic atom due to
radiative corrections. Such energy shifts break the degeneracy between
states of with the same principal quantum number 
and total angular momentum, but differing orbital and spin angular momenta.
The best known example is the energy shift between the $2S_{1/2}$ and 
$2P_{1/2}$ states in a Hydrogen-like atom,  which arises due to 
interactions of the electron with the quantum-field-theoretic
fluctuations of the electromagnetic field.  
For metric theories, the lowest order
contribution for the Lamb shift is $1052$ MHz for hydrogen
atoms. There is a $5$ MHz discrepancy with the experimental value of\
$1057.845(9)$ MHz \cite{explamb1} or $ 1057.851(2)$ MHz \cite{explamb2},
that can be improved with the inclusion of 
higher order terms and corrections coming from
the structure and recoil of the nucleus. 

Any breakdown of LPI/LLI is determined entirely by the form of the 
couplings of the gravitational field to matter since local, 
nongravitational test experiments simply respond to their external 
gravitational environment.
To explore such effects it is necesssary to develop a formalism capable
of representing such couplings for as wide a class of gravitational
theories as possible. We consider in this paper Lagrangian-based theories 
in which the dynamical equations governing the evolution of the 
gravitational and matter fields can be derived
from the action principle
\be
\label{action0}
\delta\int d^4x\,{\cal L}\equiv\delta\int d^4x({\cal L}_G+{\cal L}_{NG})=0  \quad .
\ee
The gravitational part ${\cal L}_G$ of the Lagrangian density contains only
gravitational fields; it determines the dynamics of the free gravitational
field. The nongravitational part ${\cal L}_{NG}$ contains both gravitational
and matter fields and defines the couplings between them. The dynamics
of matter in an external gravitational field follow from the action
principle
\be\label{action1}
\delta\int d^4x\,{\cal L}_{NG}=0
\ee
by varying all matter fields in an external gravitational environment. 

We  work in the context of a wide class of non-metric theories of
gravity as  described by the \tmu formalism \cite{tmu}.
Phenomenological models of ${\cal L}_{NG}$ provide a general framework for
exploring the range of possible couplings of the gravitational field to
matter and, thus, the range of mechanisms that might conceivably break LPI
or LLI. The \tmu formalism is one such model. It deals with the dynamics of
charged particles and electromagnetic fields in a static,  spherically
symmetric gravitational field. In addition to all metric theories of
gravitation, the \tmu formalism encompasses a wide
class of non-metric theories.

A quantum-mechanical extension of the original classical \tmu formalism was
developed by Will \cite{will} to calculate the energy shifts (due to {\it
e.g.} hyperfine effects) in hydrogenic atoms at rest in a \tmu
gravitational field. Since the ticking rate of a hydrogen-maser clock is
governed by the transition between a pair of these atomic states, this extension 
can be used to determine the effect of the gravitational field on the
ticking  rate of such clocks. This provides a basis for a quantitative
interpretation of gravitational redshift experiments which employ 
hydrogen-maser clocks, for example, the gravity probe A rocket-redshift
experiment \cite{redshift}.  Such experiments are a direct test of LPI.

This formalism was further extended by  Gabriel and Haugan \cite{gabriel}
who calculated  the effects the motion of an atomic system through a
gravitational field would have on the ticking rate of hydrogen-maser and
other atomic clocks. Their extension can be used to compute energies of
hyperfine and other energy shifts
of hydrogen atoms in motion through a \tmu field.
Here the physical effect under consideration is time dilation rather than
the gravitational redshift. When LLI is broken, the rates of clocks of
different types that move together through the gravitational field are
slowed by different time-dilation factors. This nonuniversal behavior is
a characteristic symptom of the breakdown of LLI \cite{haugan},
just as nonuniversal gravitational redshift is the hallmark
of LPI violation \cite{will}.

We are concerned in this paper with extending this analysis to the Lamb shift,
an energy shift whose origin is due to radiative corrections. We compute the
Gravitationally Modified (GM) Lamb Shift in a \tmu field, and then discuss
experiments which could potentially measure such effects.
We find both EEP-violating contributions to the Lamb shift
from the semiclassical \tmu Hamiltonian and its radiative
corrections. The semiclassical contribution violates
LLI only and is isotropic; the radiative corrections violate both
LLI and LPI and are not isotropic.  These contributions are functions of
non-metric parameters which arise in the leptonic sector of the standard
model, and so are not constrained by previous high-precision experiments
which have set stringent bounds for analogous parameters in the baryonic
sector \cite{PLC}. Of course all such contributions
vanish for metric theories.

In order to calculate  the (GM) radiative corrections, we shall modify
the Feynman rules of Quantum Electrodynamics
(QED) within the context of the \tmu formalism. Although
we cannot use LPI/LLI symmetries, the gauge invariance
of the theory is still present.
We shall be concerned with the one photon  contribution to the 
(GM) Lamb shift up to order $m\al(Z\al)^4$, with the nucleus treated 
as a fixed point charge.  We do not
include further (higher-order) refinements, since  we are interested in the 
role of Lamb shift energies in the investigation of possible 
LPI/LLI violations and so expect any such violations to be 
qualitatively different from higher order corrections.

Our paper is organized as follows. In Sec. II the \tmu action is introduced
and extended to frames moving with respect to the preferred frame defined by
the \tmu gravitational field. This formalism is then used
to calculate the electromagnetic fields
produced by a point-like charge and to formulate (GM)QED.
In Sec. III the (GM) Dirac equation is used to
find the  energy levels of hydrogenic atoms, and we compute the radiative
corrections for those states in Sec. IV. In Sec. V the GM
Lamb shift is related to redshift and time dilation parameters to study
possible LPI and LLI violations respectively. Final conclusions 
are presented in Sec. VI. Several appendices summarize details of our 
calculations.

\section{ (GM) Action}

The \tmu  formalism was constructed to study electromagnetically interacting
charged structureless test particles in an 
external, static, spherically symmetric (SSS) gravitational
field, encompassing a wide class of non-metric (and all metric)
gravitational theories.
 Originally employed as a
computational framework designed to test Schiff's conjecture
\cite{will1}, it permits one to extract quantitative information about
the implications of EEP-violation that can be compared to
experiment. It assumes that the non-gravitational laws
of physics can be derived from an action:
\begin{eqnarray}\label{1}
 S_{NG}&=&-\sum_a m_a\int dt\, (T-Hv_a^2)^{1/2}+\sum e_a \int dt\, v_a^\mu 
 A_\mu(x_a^\nu)\nonumber\\
& &+ \half\int d^4x\,(\epsilon E^2- B^2/\mu),
\end{eqnarray}
where $m_a$, $e_a$, and $x_a^\mu(t)$ are the rest mass, charge, and
world line of particle $a$, $x^0\equiv t$, $v_a^\mu\equiv dx_a^\mu/dt$, 
$\vec E\equiv-\vec\gr A_0-\pd\vec A/\pd t$, 
$\vec B \equiv \vec\gr\times\vec A$.
The parameters $\epsilon$, and $\mu$ are
arbitrary functions of the Newtonian gravitational potential
$U= GM/r$ (which approaches unity as $U\to 0$), as are $T$ and $H$
which in general will depend upon the species of particles within the
system  (leptons in the present case).

A quantum mechanical extension of the action (\ref{1}) which
incorporates the Dirac Lagrangian was used
by Will \cite{will} to study the energy levels of hydrogen atoms. In that
case a local approximation to the action is employed. The
spacetime scale of atomic systems allows one to ignore the spatial variations
of $T$, $H$, $\epsilon$, $\mu$, and evaluate them at the center of mass 
position of the system, $\vec X=0$. This work was further extended
by Gabriel and Haugan \cite{gabriel}
who showed that after rescaling coordinates, charges, 
and electromagnetic potentials, the field theoretic extension of the
action (\ref{1}) can be written in the form
\begin{equation}\label{act}
S=\int d^4x\, \sp(i\nn\pd+e\nn\! A-m)\psi
 + \half\int d^4x\,(E^2-c^2B^2),
\end{equation}
where local natural units are used, $\nn\! A=\gamma_\mu
A^\mu$, and $c^2=H_0/T_0\epsilon_0\mu_0$ with
the subindex ``0'' denoting the functions evaluated at $\vec X=0$.
The parameter $c$ is the ratio of the local speed of light to the
limiting speed of the species of massive particle under consideration.

The action (\ref{1}) (or (\ref{act})) has been widely used 
in the study of LPI/LLI violating
effects such as the effect of non-metric gravitational
fields on the differential ticking rates of different 
types of atomic clocks, a violation of LPI  \cite{will}.
An analysis of the electrostatic structure of atoms and nuclei
in motion through a \tmu  gravitational field using (\ref{1})
shows that the non-metric couplings 
encompassed by the \tmu  formalism can also break LLI \cite{haugan}. 
This symmetry is broken when the local speed of light 
$c_*\equiv (\mu_0\epsilon_0)^{-1/2}$
differs from the limiting speed of a given species of massive particle
$c_0\equiv (T_0/H_0)^{1/2}$,
the latter being normalized to unity in (\ref{act}).
Further implications of the breakdown of LLI on various aspects of atomic
and nuclear structure have also been investigated. Shifts in energy levels 
(including the hyperfine splitting) of hydrogenic atoms  in motion through
a \tmu  gravitational field have been calculated \cite{gabriel} 
by transforming the
representation of the action (\ref{act}) to a local coordinate system in
which the atom is initially at rest and then analyzing the atom's structure
in that frame. The local coordinate system in which the \tmu  action is
represented by Eq. (\ref{act}), is called the preferred frame; moving
frames are those systems of local coordinates that move relative  to the
preferred frame.

In the present work we generalize this analysis by using the action
(\ref{act}) to study radiative corrections to bound state energy levels in
hydrogenic atoms.  We follow the scheme given in Ref. \cite{gabriel}, and
analyze the atomic states in moving frames whose velocity is $\vec u$.

Consider an atom that moves with velocity $\vec u$ relative to the
preferred frame.  The moving frame in which this atom is initially at rest
is defined by means of a standard Lorentz transformation. 
A convenient representation \cite{gabriel} of the \tmu action in this new
coordinate system if the nongravitational fields $\psi$, $\vec A$, $\vec
E$, and $\vec B$ transform via the corresponding Lorentz transformations
laws for Dirac, vector, and electromagnetic fields is, to
$ O(\vec u^2)$,
\begin{eqnarray}\label{3}
S&=&\int d^4x\, \sp(i\nn\pd+e\nn\! A-m)\psi + \int d^4x\, J_\mu A^\mu
\nonumber\\
 &+&\half\int d^4x\,\left[(E^2-B^2)\right.\\
 &+& \left.\xi\left(\vec u^2 E^2-(\vec u\cdot\vec E)^2+(1+\vec u^2)B^2-(\vec u\cdot\vec B)^2
  +2\vec u\cdot(\vec E\times\vec B)\right)\right]\nonumber.
\end{eqnarray}
where  $J^\mu$ is the electromagnetic 4-current
associated with some external source (taken here to be a pointlike spinless
nucleus). 
In our formulation, 
all non metric effects arise from the inequality between 
$c_0$ and $c_*$ in the electromagnetic sector of the action.
The dimensionless parameter $\xi\equiv 1-(c_*/c_0)^2 =1-c^2 $ 
measures the degree to
which LPI/LLI is broken for a given species of particle.
The natural scale for $\xi$ in theories
that break local Lorentz invariance is set by the magnitude of the
dimensionless Newtonian potential, which empirically 
is much smaller than unity
in places we can imagine performing experiments \cite{will1}. 
We are therefore able
to compute effects of the terms in Eq. (\ref{3}) that break local 
Lorentz invariance via a perturbative analysis about the familiar and
well-behaved $c\to 1$ or $\xi\to 0$ limit.

The fermion sector of the action (\ref{3}) implies that the equation
of motion for the $\psi$ field is simply the Dirac equation coupled
in the usual fashion to the potential $A_\mu$. On the other hand,
the pure electromagnetic part of the action is modified with an
extra term proportional to the small (species-dependent)
parameter $\xi$. This will affect
 the electromagnetic field equations,  and  the photon propagator. In
both cases we can calculate effects of the additional terms perturbatively.

The field equations coming from the action (\ref{3}) are \cite{gabriel}
\begin{eqnarray}\label{4}
\vec\gr\cdot\vec E&=&\rho+\xi\left[\vec u\cdot\vec\gr(\vec u\cdot\vec E)-\vec
u\cdot\vec\gr\times
\vec B-\vec u^2\vec\gr\cdot\vec E\right],\\
\vec\gr\times\vec B-\dot{\vec E}&=&\vec j+\xi\left[\vec\gr\times(\vec u\times \vec E)+
\vec u\times\vec\gr(\vec u\cdot \vec B)+(1+\vec u^2)\vec\gr\times \vec B\right.\nonumber\\
&+&\left.\vec u^2\dot{\vec E}-\vec u(\vec u\cdot\dot{\vec E})
-\vec u\times\dot{\vec B}\right]\nonumber
\end{eqnarray}
where $\rho$ and $\vec j$ are the charge density and current associated
with the fermion field plus and external source (such as a nucleus.)
Perturbatively solving these equations for
electromagnetic potentials produced by a pointlike nucleus of charge $Ze$ at rest in
the moving frame yields
\begin{eqnarray}\label{po}
A_0 &=& [1-\frac{\xi}{2}(\vec u^2+(\vec u\cdot\hat n)^2)]\phi\equiv\phi+\xi\phi'
\nonumber\\
\vec A&=&\frac{\xi}{2}[\vec u+\hat n(\vec u\cdot\hat n)]\phi\equiv\xi\vec A\,'
\end{eqnarray}
where $ \hat n=\vec x/|\vec x|$, $\phi=Ze/4\pi|\vec x|$, and 
$\vec\gr\cdot \vec A=0$. Note that Eq. (\ref{po}) agrees with the corresponding
result from Ref. \cite{gabriel}. 

The primed fields in Eq. (\ref{po}) signal a breakdown of LLI.
Consequently we expect that this electromagnetic potential will modify the energy
 states of hydrogenic atoms prior to the inclusion of radiative corrections.
We shall calculate these effects for the Lamb shift in the next section.
In order to find the radiative corrections to these energy levels 
we must reformulate Quantum Electrodynamics (QED) according to the action 
(\ref{3}). Since the fermion sector of the action does not change, 
the fermion propagator is unaltered; only the photon propagator needs
to be modified.

  To find the photon propagator, we go back to the action (\ref{act}) and add 
a gauge fixing term of the form
\be\label{6}
 S_{GF}=-\half\int d^4x\,\left[(1-\xi)(\pd\cdot A)^2-2\xi \pd^0A_0\pd\cdot A\right],
\ee
after which the resulting electromagnetic part can be written as
\begin{eqnarray}\label{7}
S_{EM}&=&\int d^4x\,\left[\half A_\mu\pd^\nu\pd_\nu A^\mu\right.\\
      &+&\left.\frac{\xi}{2}(A_\mu\pd_0\pd^0 A^\mu+A_0\pd^\mu\pd_\mu A^0
-A_\mu\pd^\nu\pd_\nu A^\mu)\right]\nonumber
\end{eqnarray}
where we have integrated by parts and neglected surface terms.

This action is still given in preferred frame coordinates. We can go the
moving frame by performing the Lorentz transformations 
\begin{eqnarray}\label{8} A_0\to A_0'&=&\gamma(A_0-\vec u\cdot \vec
A)\equiv\gamma\beta\cdot A\\ \pd_0\to\pd_0'&=&\gamma(\pd_0-\vec
u\cdot\vec\gr)\equiv\gamma\beta\cdot \pd\nonumber \end{eqnarray} where
$\gamma^2\equiv 1/(1-\vec u^2)$ 
and  $\beta^\mu\equiv(1,\vec u)$;
henceforth $\beta^2\equiv 1-\vec u^2$.
Transforming Eq. (\ref{7}) by using Eq.(\ref{8}) gives
\be\label{9}
S_{EM}=\half\int d^4x\,A^\mu{\cal K}_{\mu\nu}A^\nu
\ee
where (in momentum space)
\begin{eqnarray}\label{10}
{\cal K}_{\mu\nu}=&-&\eta_{\mu\nu}k^2(1-\xi)\\
               &-&\xi\gamma^2\left[\eta_{\mu\nu}(\beta\cdot k)^2+
\beta_\mu\beta_\nu k^2\right]\nonumber
\end{eqnarray}
where $\eta_{\mu\nu}$ is the Minkowski tensor with a signature (+ - - -)
and ${\cal K}_{\mu\nu}$ is the inverse of the photon propagator
$G_{\mu\nu}$. Therefore after solving 
\be 
{\cal K}_{\mu\delta}G^{\delta\nu}=\delta_{\mu}^{\nu}, 
\ee 
we find up to first order in $\xi$ 
\begin{equation}\label{11} 
G_{\mu\nu}=-(1+\xi)\frac{\eta_{\mu\nu}}{k^2}
+\xi\frac{\gamma^2}{k^2}\left[\eta_{\mu\nu}\frac{(\beta\cdot k)^2}{k^2}
+\beta_\mu\beta_\nu\right] \quad .
\end{equation} 
The terms proportional to $\xi$ in Eq. (\ref{11}) signal the breakdown of
both LPI and LLI, since those terms are still present even  if $\vec u=0$.
The (GM) QED then differs from standard QED only in the expression for the
photon propagator; the fermion propagator and Feynman rules are
unchanged.

As the Lamb Shift is the shift between the $2S_{1/2}$ and
$2P_{1/2}$ states, and since the Dirac equation for a Coulomb
potential predicts  those states to be degenerate, the difference
between them in metric theories comes only from radiative
corrections. For non-metric theories which can be described by
the \tmu formalism these energy levels will be modified by the
EEP-violating terms introduced in the source (Eq. (\ref{po})),
removing this degeneracy before introducing radiative
corrections. Note that the fermion sector of the \tmu action does
not change and therefore neither does the Dirac equation. The
preferred frame effects appear only in the expression for the
electromagnetic source produced by the nucleus. We shall now
evaluate this contribution. 

\section{(GM) Dirac States}

 The Dirac equation in the presence of an external electromagnetic field still
reads as in the metric case:
\be \label{diraceq}
H|n\rangle=(\vec\alpha\cdot\vec p+\beta m-e A^0+e\vec\alpha\cdot\vec A)|n\rangle
=E_n|n\rangle
\ee
where the various symbols have their usual meaning.

 The (GM) energy levels of hydrogenic atoms are found by solving 
(\ref{diraceq}) in the presence of the electromagnetic field (\ref{po})
produced by the nucleus 
which entirely accounts for the  preferred frame effects. 
If we replace Eq. (\ref{po}) in  (\ref{diraceq}), the Hamiltonian 
can be written as
\be\label{eqq}
H=H_0+\xi  H',\qquad H'=-e\phi'+e\vec\alpha\cdot\vec A'
\ee
where $H_0$ corresponds to the standard Hamiltonian (with Coulomb potential only),
and  the primed fields are defined as in Eq.  (\ref{po}).
 In terms of the known solutions for $H_0|n\rangle^0=E_n^0|n\rangle^0$, we can 
perturbatively solve Eq. (\ref{diraceq}) by writing
\be\label{solution}
E_n=E_n^0+\xi E_n'\qquad |n\rangle=|n\rangle^0+\xi|n\rangle'
\ee
with
\ba\label{E'}
E_n'&=&^0\langle n |H'|n\rangle^0\equiv{E'_n}^{(E)}+{E'_n}^{(M)}\\\label{n'}
|n\rangle'&=&\sum_{r\not=n}\frac{^0\langle r |H'|n\rangle^0}{E_n^0-E_r^0}|r\rangle^0
\ea
where ${E'_n}^{(E)}$ and ${E'_n}^{(M)}$ account for the contributions
coming from the respective electric and magnetic potentials.

We now proceed to calculate the energy levels related to the 
Lamb shift states. To obtain these, we find it convenient to use the 
exact solution for the Dirac spinor  
$|n\rangle^0$, expanding the final answer in 
powers of $Z\al$ to $ O((Z\al)^4)$.  
The relationship between this approach and an alternate one in 
which the Hamiltonian  
is first expanded in powers of $Z\al$ using a 
Foldy-Wouthuysen transformation is discussed
in appendix \ref{Adirac}.

The unperturbed Dirac state $|n\rangle^0$  can be expressed as:
\be\label{spinor}
|n\rangle^0=\left (\begin{array}{r}
 G_{lj}(r)\:|l;jm\rangle\\
-iF_{lj}(r)\:\vec\sigma\cdot\nv\:|l;jm\rangle
\end{array}\right)\ee
where $|l;jm\rangle$ is the spinor harmonic 
eigenstate of $J^2$,$L^2$ and $J_z$,
with respective quantum numbers $j,l$ and  $m$.
The functions $F$ and $G$  can be written in terms of confluent
hypergeometric functions that depend in a non-trivial 
way on $Z\al$ for a given $l$ and $j$ \cite{exdirac}. 

Inserting the fields from (\ref{po}) and (\ref{spinor}) in $E_n'$, we 
write
\ba\label{coul}
{E'_n}^{(E)}&
=&\left(R_{GG}+R_{FF}\right)\langle jm;l|u^2+(\vec u\cdot\nv)^2| l; jm\rangle
\\\label{mag}
{E'_n}^{(M)}&=&-iR_{GF}\langle jm;l|(\vec\sigma\cdot\nv)(\sigma\cdot\vec u)
+\vec u\cdot\nv|l;jm\rangle
+\hbox{h.c.}
\ea
where ``h.c.'' means Hermitian conjugate and where
\ba \label{rggo}
R_{GG}=\half\int G\frac{Z\alpha}{r}Gr^2dr
\ea
with $R_{FF}$ and $R_{GF}$ defined in an analogous manner.

We now evaluate this energy for
the $2S_{1/2}$ and $2P_{1/2}$ states in this semiclassical
approximation, prior to the inclusion of any radiative corrections.
 Since the angular operator in (\ref{mag}) has odd parity (as given by $\nv$), 
it is straightforward to show that the magnetic 
contribution ${E'_n}^{(M)}=0$, so $E_n'={E'_n}^{(E)}$ for any state. 
Using the corresponding expressions for the harmonic spinors 
and the $F$, $G$ functions in (\ref{coul}) for each Lamb state
\cite{exdirac}, we find 
\ba\label{elevels}
E'_{2S_{1/2}}&=&\frac{1}{6}u^2m(Z\alpha)^2\left[1+(\frac{7}{16}+\frac{19}{16})(Z\alpha)^2
\right] +\cdots\\
E'_{2P_{1/2}}&=&\frac{1}{6}u^2m(Z\alpha)^2\left[1+(\frac{7}{16}+\frac{3}{16})(Z\alpha)^2
\right]+\cdots
\ea
where we have expanded the exact solutions for $R_{GG}$ and $R_{FF}$ 
in powers of $(Z\alpha)^2$, and kept the first relativistic correction only.
The angular integration and the $R_{GG}$  term are the same for
both states, 
and so the non-relativistic limit is still degenerate for them. However
the  first relativistic correction coming 
from the $R_{FF}$ factor breaks the degeneracy, yielding
\be\label{ED}
\Delta E_L^{(D)}=E_{2S_{1/2}}-E_{2P_{1/2}}=\xi\frac{u^2}{6}m(Z\alpha)^4+
O\left((Z\alpha)^6\right)
\ee

We obtain the result that the $2S_{1/2}$--$2P_{1/2}$ degeneracy is lifted
before radiative corrections are introduced.   
This `semiclassical' nonmetric contribution to the Lamb shift
is isotropic in the 3-velocity $\vec u$ of the moving frame and vanishes
when $\vec{u} = 0$. Hence it violates LLI but not LPI.

In order to proceed to a computation of the relevant radiative corrections,
we need to find the perturbative corrections for the energies and spinor
states given by (\ref{E'}) and (\ref{n'}) respectively. 
The radiative correction $\delta E_n$ to the Dirac energy $E_n$
can be formally expressed as
\be\label{den}
\delta E_n=\langle n|\delta H|n\rangle
\ee
where $\delta H$ accounts for the loop contributions as given by the
gravitationally modified QED. 
Since EEP violating effects appear in both the photon propagator  and the classical
electromagnetic field, we expect
\be\label{h'}
\delta H=\delta H^0+\xi\delta H'
\ee
In addition, the state $|n\rangle$ may be analogously expanded.
Up to first order in $\xi$, we can therefore write (\ref{den}) in the form
\be\label{den1}
\delta E_n=^0\langle n|\delta H^0|n\rangle^0+\xi\left[\,^0\langle n|\delta H'|n\rangle^0
+\{^0\langle n|\delta H^0|n\rangle'+\hbox{h.c.}\}\right]
\ee

The contributions from the $|n\rangle'$ states are of
the same order of magnitude (in terms of powers of $Z\al$)
as the $\delta H'$ terms and so cannot be neglected. This may be seen
by noting that, apart from the $\vec u$ dependance, 
$\phi'\sim\phi$ and so
$^0\langle n|H'|r\rangle^0\sim E^0_n-E^0_r$. Inserting this in (\ref{n'})
proves the statement.
Note that the effect of the $|n\rangle'$ 
states was overlooked in Ref. \cite{gabriel}.
If we identify $\delta H\rightarrow H_{(hf)}$,where $H_{(hf)}$
represents the perturbation to the Dirac Hamiltonian due to the 
spin of the nucleus, then by the
same arguments as before, we can show that the term
$\{^0\langle n| {H_{(hf)}}^0|n\rangle'+\hbox{h.c.}\}$ was omitted
in the corresponding expression for the hyperfine energy.

\section{ (GM) Radiative Corrections}

To lowest order in QED there are two types of radiative corrections to the
energy levels of an electron bound in an external electromagnetic
potential: the vacuum polarization ($\Pi$) and self-energy ($\Sigma$),
along with a counterterm ($\delta C$) that subtracts the
analogous processes for a free electron. These contributions are illustrated
in  Fig.1.
\begin{figure}[ht]
\centering
\leavevmode
\epsfbox[63 136 375 264] {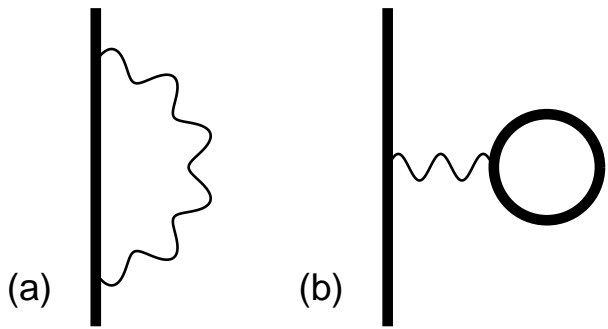}
\caption{Radiative corrections of order $\alpha$ : (a) self-energy
 and (b) vacuum polarization.} 
\end{figure}

The energy shift due to these contributions for the state $|n\rangle$ 
can then be written as
\be\label{a1}
 \delta E_n=\delta E_S+\delta E_P
\ee
where 
\be\label{as}
\delta E_S=\langle n|\Sigma-\delta C|n\rangle ,
\ee
which corresponds to the self-energy contribution in Fig. 1(a) minus the corresponding
counterterm, and
\be\label{app}
\delta E_P=\langle  n|\Pi|n\rangle ,
\ee
which is the vacuum polarization contribution
illustrated in Fig. 1(b). 

In Fig. 1 the bold line represents the bound electron propagator. This propagator can be
written in operator form as $(\nn p-\nn V-m)^{-1}$, with
$$
V^\mu(\vec x)\equiv-eA^\mu(\vec x) \quad\mbox{\rm and}\quad 
p^\mu\equiv(E_n, \vec p)
$$
where $A^\mu$ is the external electromagnetic potential. Here
$E_n$ is the total energy of the state $|n\rangle $, which satisfies the
Dirac equation
$(\nn p-\nn V-m)|n\rangle =0$

Eq. (\ref{a1}) represents the one loop correction (one power of $\alpha$)
to the atomic energy levels as given by $E_n$. We are interested in obtaining 
the ``lowest order" Lamb shift, which is the $\al(Z\al)^4$ contribution.
(There are still more approximations that come after expanding the bound
propagator, which introduce additional nonanalytic terms in the expression
for the Lamb shift that behave like $\al(Z\al)^4\ln (Z\al)$).

The GM radiative corrections are found by  evaluating (\ref{a1}) where the
external electromagnetic potential and the photon propagator are respectively
given by Eqs. (\ref{po}) and (\ref{11}). All expressions will
be expanded in terms of
the LPI/LLI violating parameter $\xi$, and the velocity of the moving frame
$\vec u$  up to $O(\xi)$ and $O(\vec u^2)$ as implied by
(\ref{po}) and (\ref{11}). EEP-violating effects are
all contained in the terms proportional to these quantities.

A variety of methods are available for evaluating the corrections in
(\ref{a1}), each differing primarily in the manner in which the bound
electron propagator is treated.  We shall follow the  method of Baranger,
Bethe and Feynman \cite{BBF} (hereafter referred to as BBF), 
in which the corrections in (\ref{as}) are
separated into a term in which the external potential acts only once, and
another term in which it acts at least twice. This latter 
`many-potential' term can be further  separated into a nonrelativistic
part, and a relativistic part which can be calculated by considering the
intermediate states as free. This approach is sufficient for the lowest
order calculation we consider here. We now proceed to outline the
main steps of this method.

The self-energy term in Eq. (\ref{a1}) can be written as
\begin{eqnarray}\label{a4} 
\delta E_S&=&\frac{\alpha}{4\pi^3}\int d^4k\;iG^{\mu\nu}(k)
\langle  n|\gamma_\mu\frac{1}{\nn p-\nn V -\nn k-m}\gamma_\nu|n\rangle -
\langle  n|\delta C|n\rangle \quad .
\end{eqnarray}
This expression gives a complex result for the level shift, since the denominators
in the integral each have a small positive imaginary part. The resulting imaginary
part of $\delta E_S$ represents the decay rate of the state $|n\rangle$ through
photon emission. The Lamb shift refers to the real part of the shift, and only 
that part will be retained in the computation of Eq. (\ref{a4}).

The difficulty in evaluating Eq. (\ref{a4}) arises entirely from choosing a convenient
expression for the bound propagator. The integrand in  
(\ref{a4}) is rearranged  in order to obtain one part which is of first order in the
potential ($\delta E_1$), and another part ($\delta E_2$) which contains 
the potential at least twice. Using the identity\cite{BBF}
\be
\hat O\equiv(\nn p_b-m)\frac{\nn p_b \hat O+\hat O\nn p_a}{p_b^2-p_a^2}-
\frac{\nn p_b \hat O+\hat O\nn p_a}{p_b^2-p_a^2}(\nn p_a-m),
\ee
to re-express $\gamma_\mu$ and $\gamma_\nu$ in (\ref{a4}) 
and respectively identifying $p_b=p$, $p_a=p-k$, and $p_b=p-k$, $p_a=p$
yields after some manipulation 
\be\label{a5}
\delta E_S=\delta E_1+\delta E_2,
\ee
where
\be\label{a6}
\delta E_1=\api\int d^3p\,d^3p'\sp_n(\vec p\,')\{I_1+I_2+I_3\}\psi_n(\vec p),
\ee
with

\ba\nonumber\label{a7}
I_1&=&\frac{i}{4\pi^2}\int\frac{2p'_\mu-\gamma_\mu\nn k}{k^2-2p'\cdot k}
\nn V\frac{2p_\nu-\nn k \gamma_\nu}{k^2-2p\cdot k}G^{\mu\nu}(k)d^4k\\
 I_2&=&\frac{i}{4\pi^2}\nn V\int\frac{2p_\mu-\gamma_\mu\nn k}{k^2-2p\cdot k}
\frac{2p_\nu-\nn k \gamma_\nu}{k^2-2p\cdot k}G^{\mu\nu}(k)d^4k\\\nonumber
I_3&=&\frac{i}{4\pi^2}\int\frac{2p_\mu-\gamma_\mu\nn k}{k^2-2p\cdot k}
 \gamma_\nu G^{\mu\nu}(k)d^4k-\delta C
\ea
and where
\ba\label{a8}\nonumber
\delta E_2&=&\frac{\alpha}{4\pi^3}\int \sp_n(\vec p\,')M_\mu(p',p'-s'-k)\\
 &\times&K_+^V(E_0-k_0;\vec p\,'-\vec s\,'-\vec k,\vec p+\vec s-\vec k)\\
 &\times&M_\nu^{\dag}(p+s-k,p)\psi_n(\vec p)G^{\mu\nu}(k)d^4k\,d^3p\,d^3p'd^3s\,d^3s'
 \nonumber\\
 &\equiv& \langle  M K_+^V M \rangle\nonumber
\ea
with
\ba\label{a9}
M_\mu(p',p-k)&=&\nn V(\vec p\,'-\vec p)\frac{2p_\mu-\gamma_\mu\nn k}
                  {2p\cdot k-k^2}
-\frac{2p_\mu'-\gamma_\mu\nn k}{2p'\cdot k-k^2}\nn V(\vec p\,'-\vec p)\nonumber\\
M_\nu^{\dag}(p'-k,p)&=&\nn V(\vec p\,'-\vec p)\frac{2p_\nu-\nn k\gamma_\nu}
                  {k^2-2p\cdot k}
-\frac{2p_\nu'-\nn k \gamma_\nu}{k^2-2p'\cdot k}\nn V(\vec p\,'-\vec p)\nonumber
\ea
The quantity $K_+^V$ is defined as $ -iK_+^V\equiv(\nn p-\nn V-m)^{-1}$,where in
 momentum space $K_+^{V}=\delta(E'-E)K_+^V(E;\vec p\,',\vec p)$.

In Eqs.(\ref{a6}) and (\ref{a8}) the $p$'s have time component $E_n$  and
the $s$'s have time component 0. Note that the above derivations are
independent of the specific form of the photon propagator $G_{\mu\nu}$.

Further evaluation entails a lengthy computation which in principle is
analogous to that of BBF. In practice though, the calculation is 
substantially more complicated than in the metric case due to the 
additional non metric
terms present in the photon propagator and the electromagnetic source
related to a charged point particle. Regularization and renormalization
procedures have to be modified accordingly as well.
Details involving  the subsequent computation of the self energy  (and
vacuum polarization) term are given in appendix \ref{Aloop}. 

The final result for the loop corrections related to the Lamb shift is
of the form  
\be\label{EQ}
\Delta E^{(Q)}_L=\delta E_{2S_{1/2}}-\delta E_{2P_{1/2}}
\ee
where each term is obtained from  eq. (\ref{efinal}) (and its
relevant subsidiary equations) as calculated for the corresponding 
atomic state. By adding 
the ``semiclasical" correction
coming from the Dirac level (labeled by $(D)$ in Sec. III), 
the total Lamb shift reads
\ba\label{EDQ}\nonumber
\Delta E_L&=&\Delta E^{(D)}_L+\Delta E^{(Q)}_L\\
&=&\frac{m}{6\pi}(Z\al)^4\al\Big\{-2.084+\ln{1\over\al^2}
+\xi\Big[-4.534+\frac{3}{2}\ln{1\over\al^2} \\\nonumber
&+&\vec u^2\left[\frac{\pi}{\al}
-3.486+\frac{2}{3}\ln{1\over\al^2}-0.011\,\cos^2\theta\right]+ u_iu_j\Delta\hat\epsilon^{ij}\Big]\Big\}
\ea
where  we have introduced the dimensionless parameter $\Delta\hat\epsilon^{ij}\equiv
2\Delta\hat E^{ij}/((Z\al)^4m^3)$ (see (\ref{Eij})),   and used Eqs. (\ref{refener}) and
(\ref{E0}) in the evaluation of (\ref{EQ}) through Eq. (\ref{efinal}).

The former result is the energy shift associated with the particular 
states in (\ref{EQ}). However in Eq. (\ref{efinal}) we have derived a 
general expression for the one-loop radiative corrections related to any 
atomic  state. These are
\be\label{en0}
\delta E_{n0}=\frac{4}{3\pi}\frac{(Z\al)^4\al}{n^3}m\left[\frac{19}{30}
-\frac{\xi}{30}+(1+\frac{3}{2}\xi)\ln(\frac{m}{2E_*^{n0}})+O(u^2)\right]
\ee
for $l=0$, and
\be\label{enl}
\delta E_{nl}=\frac{4}{3\pi}\frac{(Z\al)^4\al}{n^3}m\left[(1+\frac{3}{2}\xi)
\ln(\frac{Z^2 Ryd}{E_*^{nl}})+\frac{3}{8}\frac{C_{lj}}{2l+1}(1+\frac{\xi}{2})
+O(u^2)\right]
\ee
for $l\not=0$; where we have not explicilty written the terms proportional 
to the moving frame velocity.  Here
\be
C_{lj}=\left\{\begin{array}{ll}
1/(l+1)&\hbox{for $j=l+1/2$}\\
-1/l&\hbox{for $j=l-1/2$}
\end{array}\right.
\ee
and $E_*$ is defined by (\ref{erefer}). Values for this
 reference energy can be obtained from Ref.\cite{harriman}
up to states with $n=4$.

Note that in addition to the explicit dependence on the frame
velocity in Eq. (\ref{EDQ}), there exists a position dependence hidden
by the  rescaling of the original action (Eq. (\ref{3})), which was considered locally constant 
throughout the computation. The full \tmu parameter dependence
in Eq. (\ref{EDQ}) can be recovered by replacing
\be\label{rede}
\al\rightarrow\al\frac{1}{\epsilon}\sqrt{\frac{H}{T}},\qquad m\rightarrow
m\sqrt{H},\qquad\Delta E_L\rightarrow\sqrt{\frac{H}{T}}\Delta E_L
\ee
in the preceding equations.

Note that $\xi$ in Eq. (\ref{EDQ}) accounts for any EEP violation
coming from a non-universal gravitational coupling between photons
and leptons. A further distinction can still be made between
leptons and antileptons. In principle a matter/antimatter violation
of the EEP could be measured in a Lamb shift transition, through the
appearance of virtual positron/electron pairs in the vacuum polarization
loop contribution \cite{Schiff}.   
This will add a non metric term to Eq. (\ref{EDQ}),
of the form (see appendix \ref{Apol} for more details):
\be\label{posivac}
\Delta E_L^{(+)}=-\xi_{e_+}\frac{m}{120\pi} (Z\al)^4\al  (1+2|\vec u|^2)
\ee
where $\xi_{e_+}=1-c_{e_-}/c_{e_+}$ accounts for the difference between
the limiting speed of electrons ($c_{e_-}=c_0$) and positrons ($c_{e_+}$).

We turn next to the question of relating the Lamb shift 
to observable quantities in order to parameterize possible violations.
of the EEP.

\section{Test for LPI/LLI Violations}
 
We begin by considering a general idealized composite body made up of
structureless test particles that interact by some nongravitational force to
form a bound system. The conserved energy function of the body $E$ is
assumed to have the quasi-Newtonian form \cite{haugan}
\be\label{pp1}
E=M_Rc_0^2-M_RU(\vec X)+\half M_R|\vec{V}|^2+..
\ee
where $\vec X$ and $\vec V$ are respectively the 
quasi-Newtonian coordinates and velocity of
the center of mass of the body, $M_R$ is the rest energy of the body and $U$
is the external gravitational potential. Potential violations of the EEP
arise when the rest energy $M_R$ has the form
\be
M_Rc_0^2=M_0c_0^2-E_B(\vec X,\vec V)
\ee
where $M_0$ is the sum of the rest masses of the structureless constituent
particles and $E_B$ is the binding energy of the body. It is the position
and velocity dependence of $E_B$, which signals the breakdown of the EEP. 
Expanding $E_B$ in powers of $U$ and $V^2$ to an order consistent with
(\ref{pp1}) we have
\be\label{bin}
E_B(\vec X,\vec V)=E_B^0+\delta m_P^{ij}U^{ij}-\half\delta m_I^{ij}V^iV^j
\ee
where $ U^{ij}$ is the external gravitational potential tensor, satisfying
$U^{ii}=U$. The quantities $\delta m_P^{ij}$ and $\delta m_I^{ij}$ are
respectively called the anomalous passive gravitational and 
inertial mass tensors. 
They depend upon the detailed internal structure of the
composite body. In an atomic system they
  can be expected to consist of terms proportional to the
electrostatic, hyperfine, Lamb shift, and other contributions to the binding
energy of an atomic state.

In a gravitational redshift experiment one compares the local energies at
emission $E_{em}$ and at reception $E_{rec}$ of a photon transmitted between
observers at different points in an external gravitational field. The 
measured redshift is  defined as 
$$
Z=\frac{E_{em}-E_{rec}}{E_{em}}
$$
Using (\ref{pp1}) (with $\vec V=0$)
to relate the transition energies at the two different points, this
parameter can be expressed as \cite{haugan}
\be\label{redz}
Z=\Delta U\Big(1-\Xi\Big),
\qquad
\Xi=\frac{\delta m_P^{ij}}{\Delta E_B^0}\frac{\Delta U^{ij}}{\Delta U}
\quad .
\ee
Clearly $Z$ depends (through $\delta m_P^{ij}$) upon the 
specific test system used in
 the experiment. An absence of LPI violations 
will mean  $\Xi=0$, and so $Z$ will be
independent of the detailed physics underlying the energy transition .

The LLI violations may be empirically probed through time dilation experiments.
These experiments compare  
atomic energy transitions as measured by the moving
frame  ($\Delta E_B$) and 
preferred frame ($\Delta E_B^0$), which can be related via \cite{gabriel}
\be
\Delta E_B=\Delta E_B^0\Big(1-[A-1]\frac{\vec V ^2}{2}\Big)
\ee
with  the time dilation coefficient $A$ defined by
\be\label{timedi}
A=1-\frac{\delta m_I^{ik}}{\Delta E_B^0}\frac{V^iV^k}{V^2}
\quad .
\ee

 Here $\delta m_I^{ik}$ represents the difference between 
the anomalous inertial
tensors related to the atomic states involved in the transition.
  The coefficient $A$ represents the dilation of the rate of a 
moving atomic clock
whose frequency is governed by the transition. Since the anomalous
mass tensor is not isotropic, $A$ depends upon the orientation of the atom's
quantization axis relative to its velocity through the preferred
frame. Note that if LLI is valid the anomalous inertial mass tensor 
associated with every atomic state vanishes, so that $A=1$.

Here we consider the possibility of employing
the Lamb shift as the atomic transition governing the appropriate 
experiment. To do so we must compute the relevant
$\Xi$ and $A$ coefficients respectively.

In order to calculate the corresponding $\delta m_p^{ij}$ 
related to the Lamb shift,
we  must find the manner in which $\Delta E_L$ varies as the location of the
atom is changed. Setting $\vec u=0$ in (\ref{EDQ})
and performing the rescaling given in (\ref{rede}), we obtain
\be\label{lpi}\nonumber
\Delta E_L
={\cal E}_L\frac{\sqrt{T}}{\epsilon^5}\left(\frac{H}{T}\right)^{5/2}
\left\{1+a\xi +b\,(1+\frac{3}{2}\xi)\,\ln\left(\epsilon^2\frac{T}{H}\right)\right\}
\ee
with ${\cal E}_L=\frac{m}{6\pi}(Z\al)^4\al /b$, and
$$a=b(-4.534+\frac{3}{2}\ln{1\over\al^2})\qquad
b=1/(-2.084+\ln{1\over\al^2})$$
where $ {\cal E}_L$ represents the metric value (within the
given approximations) for the Lamb shift. Note that there is still a position
dependence in (\ref{lpi}) through the definition of
\be
\xi\equiv 1-\frac{H}{T}\frac{1}{\mu\epsilon} \quad .
\ee

We recall that the total energy of the system can be expressed in term of
\be\label{etotal}
E=m\sqrt{T}+\Delta E_L+\cdots
\ee
where $\cdots$ represents other contributions for the binding energy of the 
system.

The functions $T$, $H$, $\epsilon$ and $\mu$, considered to be functions of $U$
and evaluated at the instantaneous center of mass location $\vec X=0$ for
purposes of the calculation of $\Delta E_L$, are now expanded in the form
\be\label{tu}
T(U)=T_0+T_0'\vec g_0\cdot\vec X+O(\vec g_0\cdot\vec X)^2
\ee
where $\vec g_0=\vec\gr U|_{\vec X=0}$, $T_0=T|_{\vec X=0}$, and
$T_0'=dT/dU|_{\vec X=0}$. It is useful to redefine the gravitational
potential $U$ by
\be\label{u}
U\to-\half\frac{T_0'}{H_0}\vec g_0\cdot\vec X
\ee
whose gradient yields the test-body acceleration $\vec g$.

If the above is used to expand (\ref{etotal}), we get
\be\label{eee}
E=(m+{\cal E}_L)(1-U)+{\cal E}_L U\Big\{(5-a-2b)\Gamma_0
-a\Lambda_0\Big\}
\ee
where we have used (\ref{rede}); and
neglected terms proportional to $\xi$, since the main
position dependence parameterization is given in terms of:
\be\label{para}
\Gamma_0=\frac{2T_0}{T_0'}(\frac{\epsilon_0'}{\epsilon_0}+
\frac{T_0'}{2T_0}-\frac{H_0'}{2H_0}),
\qquad
\Lambda_0=\frac{2T_0}{T_0'}(\frac{\mu_0'}{\mu_0}+
\frac{T_0'}{2T_0}-\frac{H_0'}{2H_0})
\ee

If we now  identify (\ref{eee}) with Eqs.  (\ref{pp1}) and (\ref{bin}), 
we  can obtain the corresponding Lamb shift contributions to the binding
energy and anomalous passive mass tensor as
\ba\label{massten}
\Delta E_B^{0(L)}&=&-{\cal E}_L\\
\delta m_P^{ij(L)}&=&\Delta E_B^{0(L)}\Big\{(5-a-2b)\Gamma_0
-a\Lambda_0\Big\} \quad .\nonumber
\ea
This result was first presented in Ref. \cite{cm}, where in (\ref{massten}) we have corrected 
the latter for a sign error in the coefficient multiplying $\Lambda_0$ and
a missing factor $b$ in the $\Gamma_0$ term.

Inserting (\ref{massten}) in (\ref{redz}), we obtain 
\be
\Xi^L=3.424\,\Gamma_0-1.318\,\Lambda_0
\label{lamlpi}
\ee
as the LPI violating parameter associated with the Lamb shift transition.
Note that if LPI is valid then $\Gamma_0=\Lambda_0=0$.

In comparing the result (\ref{lamlpi}) to anomalous redshift parameters
computed for other systems, it is important to note that we are working 
with units that are species dependent.  Recall that the choice of $c_0=1$,
 and the redefinition of the gravitational potential (\ref{u})
 involves the $T$ and $H$ functions associated with electrons
(or more generally a given species of lepton). 

Consider, for example, hyperfine transitions (maser clocks). In this
case the leptonic and
baryonic gravitational parameters appear  simultaneously. This
atomic splitting comes from the interaction between the magnetic
moments of the electron and proton (nucleus). The proton metric
appears only in the latter, and so it does not affect the principal
and fine structure atomic energy levels. It is simple to check that
the hyperfine splitting scales as
\be\label{hf}
\Delta E_{hf}={\cal E}_{hf} \frac{{T_B}^{1/2}}{H_B}\frac{H_0^2}{T_0}
\frac{\mu_0}{\epsilon_0^3}
\ee
where the label $B$  is added to distinguish baryonic related functions from
leptonic ones; and ${\cal E}_{hf}$ depends only on atomic parameters.

In expanding (\ref{hf}) according to (\ref{tu}), we obtain
\be\label{hf1}
\Delta E_{hf}={\cal E}_{hf}(1-U_B)+{\cal E}_{hf} U_B\Xi^{hf}
\ee
with
\be
\Xi^{hf}=3\Gamma_B-\Lambda_B+\Delta
\ee
where $U_B$, $\Gamma_B$ and $\Lambda_B$ are the baryonic analogues of
(\ref{u}), and (\ref{para}) respectively. 
In (\ref{hf1}) we rescaled the atomic parameters to absorb the
\tmu functions and chose units such that $c_B=1$. The quantity $\Delta$ is
given by 
\be
\Delta=2\frac{T_B}{T_B'}\left[2(\frac{H_B'}{H_B}-\frac{H_0'}{H_0})
-\frac{T_B'}{T_B}+\frac{T_0'}{T_0}\right]
\ee
and would vanish under the assumption that the leptonic and baryonic
\tmu parameters were the same.

Turning next to experiments which test LLI, 
we need to obtain the tensor $\delta m_I^{ij}$ appropriate to the Lamb
shift. This tensor is obtained after taking  partial derivatives
of  $\Delta E_L$ with respect to $u_i$ and $u_j$ (note $\vec
V\equiv\vec u$). Substituting the result into (\ref{timedi}) yields
\be\label{al}   
1-A^L=\frac{\xi}{7.757}\left\{\frac{\pi}{\al}+
3.074-0.011\,\cos^2\theta+
\frac{V_iV_j}{\vec V^2}\Delta\hat \epsilon^{ij}\right\} 
\ee
for the Lamb shift time dilation coefficient, 
where $\theta$ is the angle between the atom's quantization axis and its
velocity with respect to the preferred frame.

Note that the coefficient $A_L$ depends upon $\Delta\hat
\epsilon^{ij}$, the evaluation of which
involves the computation of an infinite sum as given by (\ref{Eij}).
The dominant contribution in Eq. (\ref{al}) comes
from the Dirac part of the energy (proportional to $\frac{1}{\al}$ ), 
which produces an overall shift only.  Non-isotropic effects arise 
solely due to radiative corrections.

In general, an experimental test of LLI involves a search for the effects of motion relative
to a preferred frame such as the rest frame of the cosmic microwave
background. A detailed analysis about the interpretation of LLI 
violating experiments is presented in Ref. \cite{gabriel}, which
analyzed experiments concerned with hyperfine transitions, obtaining
an expression for the time dilation parameter corresponding to that 
kind of transition\footnotemark
\footnotetext{Note that the expression given there 
for $A^{Hf}$  is incomplete according to discussion presented in Sec. III}.
This parameter is negligible in comparison with other sources of energy, 
such as nuclear electrostatic energy in the case of the
$^9$Be$^+$ clock experiment \cite{PLC1}. 

In summary, we have been able to parameterize EEP violations arising from 
Lamb shift transitions associated with redshift and time dilation 
experiments. In these types of EEP violating experiments 
one typically looks for variations of the
energy shift due to changes in either the gravitational  potential or the
direction of the preferred frame velocity. 
The feasibility of such experiments is hindered
by the present  level of precision of Lamb shift transitions 
(one part in $10^6$) in comparison to the magnitudes of such changes.
In the first case, any Earth based
experiments will be limited by the small size of the
Earth's gravitational potential ($\approx 10^{-9}$), which is
well beyond any foreseeable  improvement in Lamb shift precision. 
Similar problems
appear in the second case, where the known upper bound
$|\vec{u}| < 10^{-3}$ \cite{will1} for the preferred frame velocity,
leaves no room for any improvement on the EEP violating
parameter $\xi$, since anisotropic effects go as  $\xi |\vec{u}|^2$. 

However useful information can still be extracted from Eq. (\ref{EDQ})
if we use the current level of discrepancy between the experimental
result \cite{explamb1} and the theoretical (metric) value \cite{Eides}
to bound the nonmetric contributions for the Lamb shift. This constrains
$\xi< 1(1)\times10^{-5}$. Similar bounds can be obtained by
considering empirical information about other atomic states.
In this context, the indirect measurement of the $1S$ Lamb shift
\cite{weitz} gives a limit $\xi< 1.4(1) \times10^{-5}$, and the
measurement of the $2S_{1/2}-2P_{3/2}$ fine structure
interval \cite{hagley}: $\xi < 0.7 (1.4)\times 10^{-5}$.
If we drop the assumption that positrons and
electrons have equivalent couplings to the gravitational field
\cite{Schiff}, we find that there is an 
additional contribution to (\ref{EDQ}) due to $\xi_{e^+} 
\neq \xi_{e^-}$. This contribution arises entirely from radiative 
corrections and is given by eq. (\ref{posivac}).  Making the same 
comparisons as above,  we find the most 
stringent bound on this quantity to be $|\xi_{e^+} | < 10^{-3}$.

The previous bounds were obtained by using  (\ref{coul}) and (\ref{en0}) 
or (\ref{enl}) to calculate the corresponding nonmetric Dirac and
radiative corrections contributions respectively. The $1S$
Lamb shift experiment, actually measures the transition:
$(E_{4S}-E_{2S})-\frac{1}{4}(E_{2S}-E_{1S})$, and so
we use this one to make the comparison, where experimental
and theoretical values are given in ref.\cite{weitz}.
In the other experiment we need to use the non metric part
of $E_{2S_{1/2}}-E_{2P_{3/2}}$ ($\equiv \Delta_\xi$), namely:
\be\label{fine}
\Delta_\xi=\xi(Z\al)^2m\left[\pm\frac{u^2}{60}(\frac{3}{2}\cos^2\theta-1)
+O((Z\al)^2u^2)+\frac{\al(Z\al)^2}{6\pi}(10.434+O(u^2))\right]
\ee
where the first term comes from the Dirac contributions (here
+ and - label the transition coming from the $2P_{3/2}$  state
with $|M|=3/2$ and $|M|=1/2$ respectively)
and the second one from radiative corrections. Note that the
leading anisotropic effects stem from the non relativistic 
contributions, and so their ratio with the metric value,
$O(m( Z\al)^4)$, is $O(\xi u^2 /(Z\al)^2)$ , instead of $O(\xi u^2)$ as for
the {\it classical} Lamb shift. Time dilation experiments will look
for changes on the $E_{2S_{1/2}}-E_{2P_{3/2}}$ splitting as the
Earth rotates, which would single out only the preferred frame
contributions. Current experiments \cite{hagley} measure a value of
$9911.200(12)$ MHz for that transition, 
which gives a nominal bound (coming from the experimental error) 
of $\frac{3}{2}\xi\cos^2\theta<1\times 10^{-4}$ for the preferred frame 
part. This bound should improve once appropriate experiments are carried 
out, since these will look for periodic behaviour which can be isolated  
and measured with high precision.

Note that  an empirical  value for the Lamb shift is obtained from
Ref.\cite{hagley}  by subtracting the theoretical result of the 
fine splitting
$2P_{1/2}-2P_{3/2}$. Now by following the previous formalism we can 
parameterize the LPI violation in the former experimental
result through:
 \be\label{fineb}
E_{2S_{1/2}-2P_{3/2}}^{exp}=({\cal E}_f+{\cal E}_L)(1-U)
+U({\cal E}_f\Xi^f+{\cal E}_L\Xi^L)
\ee
where we have added the corresponding parameters
related to the fine transition \cite{will1}: ${\cal E}_f$ and $\Xi^f$.
Constraining the ratio of this quantity to a direct measurement of 
the Lamb shift  \cite{explamb1} to lie within experimental/theoretical
error, we obtain the bound
$|U(\Xi^L-\Xi^f)|=|U(0.576\Gamma_0+1.318\Lambda_0)| < 10^{-5}$.
This result is sensitive to the
absolute value of the total local gravitational potential
\cite{Hughes,Good}, whose
magnitude has recently been estimated to be as large as
$3\times 10^{-5}$ due to the local supercluster \cite{Kenyon}. 
Hence measurements of this type can provide us with empirical
information sensitive to radiative corrections that
constrains the allowed regions of $(\Gamma_0,\Lambda_0)$ parameter
space. Unfortunately the present level of precision in measuring the 
Lamb shift allows only a rather weak constraint.

\section{Discussion}

We have computed for the first time radiative corrections to a physical
process, namely the energy shift between two hydrogenic energy levels that
are semi-classically degenerate, within the context of the \tmu formalism.
The corresponding (GM) QED was derived, and the
(GM) expressions for the propagators were obtained. The nonmetric aspects
of a theory describable by the \tmu formalism
 can be all included in the  photon propagator, given an appropriate
choice of coordinates, leaving the fermion propagator unchanged.
The addition of more parameters to the theory (by the \tmu functions)
entail new renormalizations, where not only charge and mass need to be
redefined but also the \tmu parameters.

The approach we took to solve for the semi-classical Dirac energies
(sec. III) differs from the one given in Ref. \cite{gabriel}, in which the Dirac Hamiltonian was 
expanded using Foldy-Wouthuysen transformations yielding the first relativistic
correction to the Schr\"odinger Hamiltonian (as introduced for example, for
the Darwin and spin-orbit  terms), and subsequently the energies.  Instead we began
from  the fully relativistic expression, where the perturbations
come only from the preferred frame terms of the electromagnetic
potential. Our approach involved evaluating expectation values with respect
to the relativistic spinors instead of their nonrelativistic extensions (or Pauli
states). The effects of  relativistic corrections such as spin-orbit coupling are therefore
included exactly in this approach. Once this is done, the final result
is expanded to keep it within the desired order. The semi-relativistic
approach is not suitable when preferred frame effects are studied.

Qualitatively new information on the validity of
the EEP will be obtained by  setting new empirical bounds on the
parameters $\xi$, $A_L$ and $\Xi_L$  which are associated with purely
{\it leptonic} matter. Relatively little is known about
empirical limits on EEP-violation in this sector \cite{Hughes}.
Previous experiments have set the limits
\cite{PLC} $|\xi_B| \equiv |1 -c_B^2|\,<\,6\,\times \,10^{-21}$
where $c_B$ is the ratio of the limiting speed of baryonic matter
to the speed of  light. In our case we obtain an analogous bound on
$\xi$ for electrons from the difference between current experimental
and theoretical values, giving 
 $|\xi| < 10^{-5}$. Although much weaker than the bounds on
$\xi_B$, it is comparable to that
noted in a different context by Greene {\it et. al.} \cite{Greene}.
They considered a similar formalism (\tmu with $\vec u=0$) for analyzing
the  measurement of the photon wavelength emitted in
a transition where a mass $\Delta m$ is converted into
electromagnetic radiation, thereby providing an empirical
relationship between the limiting speed of massive 
particles (electrons) and light.

The breakdown of LPI for the Lamb shift in the context of a nonmetric theory of
gravity describable by the \tmu formalism is embodied in the
the anomalous gravitational redshift parameter (\ref{lamlpi}).
Recall that $\Xi$ depends on the nature of the atomic transition
through the evaluation of the anomalous passive tensor. This tensor will
have differing expressions  for differing  types of atomic transitions
\cite{will1}. An atomic clock based on the Lamb shift transition will,
in a non-metric theory, exhibit a ticking rate that is dependent upon the
location of the spacetime frame of reference and that differs from
frequencies of clocks of differing composition. For example,
the gravity probe A experiment \cite{redshift} employed hydrogen-maser
clocks, and was able to constrain the corresponding LPI violating
parameter related to hyperfine transitions:
\begin{equation}
|\Xi^{Hf}|=|3\Gamma_B-\Lambda_B +\Delta|<2\;\times\;10^{-4}
\label{hyplpi}
\end{equation}
This experiment involves interactions between
nuclei and electrons and so does not (at least to the leading order to
which we work) probe the leptonic sector in the manner
that Lamb-shift experiments would.
In general Eq. (\ref{redz}) will describe
the gravitational redshift of a photon emitted due to
a given transition in a hydrogenic
atom; for a  hyperfine transition the redshift parameter is
(\ref{hyplpi}), whereas it is
(\ref{lamlpi}) for the Lamb shift transition.

An analogous experiment to test for LPI violations based on Lamb shift
transition energies poses a formidable experimental challenge
because of the intrinsic uncertainties of excited states of Hydrogenic
atoms. Setting empirical bounds on $\Xi_L$ 
by precisely comparing two identical Lamb shift transitions
at different points in a gravitational potential would appear unfeasible
since the anticipated redshift in the background potential of
the earth ($\approx 10^{-9}$) is
much smaller than any foreseeable improvement
in the precision of Lamb-shift transition measurements \cite{Eides}.
One would at least need to  perform the experiment in a
stronger gravitational field (such as on a satellite in close solar
orbit) with 1-2 orders-of-magnitude improvement in precision.
A `clock-comparison' type of experiment between a
`Lamb-shift clock' and some other atomic frequency standard \cite{will1}
is, in principle, sensitive to the
absolute value of the total local gravitational potential
\cite{Hughes,Good}, as noted earlier. With this interpretation,
comparitive transition measurements of the type discussed in the 
previous section can more effectively constrain the allowed regions
of $(\Gamma_0,\Lambda_0)$ parameter space than can measurements which
depend upon changes in the gravitational potential. Of course
exploiting anticipated improvements in precision of measurements 
of atomic vacuum energy shifts \cite{Eides} will yield
better bounds on $\xi_{e^-}$ and $\xi_{e^+}$ via (\ref{EDQ}).

Violations of LLI single out a preferred frame of reference. In fact,
the search for a preferred direction motivated the most precise tests of
LLI performed so far \cite{PLC1,PLC}. We have extended the analysis of
the effects of motion relative to a preferred frame to account for the
radiative correction for the atomic energies associated with the Lamb shift,
as embodied in the expression (\ref{al}). This non-universality reflects the
breakdown of spatial isotropy for quantum-mechanical vacuum energies.
The coefficient $A_L$ depends upon $\Delta\hat \epsilon^{ij}$, the
evaluation of which
involves the numerical computation of the sum in (\ref{Eij}).
Unfortunately, the intrinsic linewidths of the relevant states
render direct measurement of such effects unfeasible.  More precise 
empirical information on the value of $\xi$ can be obtained by precisely 
measuring changes in the $E_{2S_{1/2}}-E_{2P_{3/2}}$ splitting as functions
of terrestrial or solar motions.  However these effects are insensitive
to radiative corrections, depending instead upon
the semi-classical non-metric effects discussed in section III.

Finally, we note that our formalism could be applied for muonic atoms.
For a muon-proton bound system, we will obtain an expression similar
to that of (\ref{efinal}), but where all parameters refer to muons.
For an anti-muon electron bound system (a muonic atom) a similar analysis
would apply. However in both cases the mass and spin of the muon could
not be neglected.

We expect that the intrinsically quantum-mechanical character
of the radiative corrections will motivate
the development of new LPI/LLI experiments based
on the  Lamb shift transition. In so doing
we will extend our understanding of the validity of the
equivalence principle into the regime of quantum-field theory.

\vspace{3 cm}

\section*{Acknowledgements}
This work was supported in part 
by the Natural Sciences and Engineering Research
Council of Canada.  
We are grateful to C.M. Will for his initial encouragement in
this work, and to M. Haugan and R. Moore for helpful discussions.

\setcounter{equation}{0}
\setcounter{section}{-1}
\begin{center}
\section{ Appendices}
\end{center}

\renewcommand{\thesection}{\Alph{section}}
\renewcommand{\theequation}{\Alph{section}\arabic{equation}}
\section{Semi-Relativistic Calculation of Hydrogenic Energy Levels
\label{Adirac}}

Consider a hydrogenic atom immersed in an external gravitational
field,  moving with velocity $\vec u$ relative to the preferred frame.
In Sec. III we follow a fully relativistic 
approach to solve for the atomic energy
levels. That is, we perturbatively solve the Dirac equation in the presence of
the electromagnetic field of the nucleus, where the unperturbed
states correspond to the Dirac solution in the presence
of a Coulomb potential only (the metric case). 

We consider here the use of the  Foldy-Wouthuysen transformation in solving
(\ref{diraceq}).  In this approach, we write
\be
H=H_{c}+H_{mag}+H_{mv}+H_{SO}+H_{D}
\ee
with
\ba\label{semi}
H_{c}&=&m+\frac{\vec p^2}{2m}-eA_0\nonumber\\
H_{mag}&=&\frac{e}{2m}(\vec p\cdot\vec A+\vec A\cdot\vec p)+
\frac{e}{2m}\vec\sigma\cdot\vec B\nonumber\\
H_{mv}&=&-\frac{\vec p^4}{8m^3}\\
H_{SO}&=&\frac{ie}{8m^2}\vec\sigma\cdot\vec\gr\times\vec E
+\frac{e}{4m^2}\vec\sigma\cdot\vec E\times\vec p\nonumber\\
H_D&=&\frac{e}{8m^2}\vec\gr\cdot\vec E\nonumber
\ea
where $A_\mu$ is given by Eq. (\ref{po}).

As shown in section III, we can take $H_{mag}\rightarrow 0$, since 
the magnetic field does not contribute to the atomic energy levels.   
We can then group the terms in the Hamiltonian as
\ba
H&=&H_{c}+H_f\nonumber\\
H_f&=&H_{mv}+H_{SO}+H_D
\ea
where we have defined the fine contribution to the Hamiltonian ($H_f$), in order to
account for the first relativistic correction $O((Z\al)^4)$ to the atomic energy levels.

We start writting a formal solution for $H|n\rangle=E_n|n\rangle$, in term
of its non-relativistic limit:
\be\label{12}
H_{c}|n\rangle_{c}=E_n^{c}|n\rangle_{c},
\ee
as
\be\label{13}
|n \rangle=|n \rangle_{c}+|n\rangle_f,\qquad E_n=E_n^{c}+\,_{c}\langle n|H_f|n \rangle_{c}
\ee
where the index ``$f$" accounts for the first relativistic correction to the
states and energies.

Since $A_0=\phi+\xi\phi'$, and so $H_{c}=H_{c}^0+\xi H_{c}'$,
we do not know the exact solution for (\ref{12}), but the perturbative
expansion:
\be\label{17}
|n\rangle_{c}={|n\rangle_{c}}^0+\xi|n\rangle_{c}'\qquad
{E_n}^{c}={E_n}^{c(0)}+^0_{c}\langle n|H_{c}'|n \rangle_{c}^0
\ee
where 
\be\label{17a}
H_{c}^0|n\rangle_{c}^0=(m+\frac{\vec p^2}{2m}-e\phi)|n\rangle_{c}^0=
E_n^{c(0)}|n\rangle_{c}^0
\ee

If we use (\ref{17}) along with $H_f=H_f^0+\xi H_f'$ in (\ref{13}), 
 we can finally write up to $O(\xi)$,
\ba\label{222}
E_n&=&E_n^0+\xi E_n'
=\,^0_{c}\langle n|\left(H_c^0+H_f^0\right)|n \rangle_{c}^0\\
&+&\xi\left[\,^0_{c}\langle n|\left({H_c}'+{H_f}'\right)|n \rangle_{c}^0+
\left\{\,^0_{c}\langle n|H_f^0|n \rangle_{c}'+\hbox{h.c.}\right\}\right]+O((Z\al)^6)
\nonumber
\ea

We see then that under this semi-relativistic approach,
 we must address  the problem of finding the states $|n \rangle_{c}'$,
whose contribution to (\ref{222}) is between the brace brackets.
This is equivalent to include the first relativistic correction
coming after solving
\be
H^0|n\rangle^0=(H_{c}^0+H_f^0+\cdots)|n\rangle^0
\ee
as
\be
|n\rangle^0=|n\rangle^0_{c}+|n\rangle^0_f+\cdots,
\ee
since, we can show
\be\label{hc}
\left\{\,^0_{c}\langle n|H_f^0|n \rangle_{c}'+\hbox{h.c.}\right\}=\left\{\,_f^0\langle n|H_{c}'|n \rangle_{c}^0
+\hbox{h.c.}\right\}
\ee

This relation allows us to rewrite part of (\ref{222}) as
\ba\label{26}
{E_n}'&=&\left(\,_{c}^0\langle n|+\,_f^0\langle n|+\cdots\right)\left(H_{c}'+H_f'+\cdots\right)
\left(|n\rangle_{c}^0+|n\rangle_{f}^0+\cdots\right)\nonumber\\
&=&\,^0\langle n|H'|n\rangle^0.
\ea

It is clear then that if we start with the exact solution  for
the Dirac equation in the presence of a Coulomb potential, 
we can avoid working with the states $|n \rangle_{c}'$. Note that since 
we are interested only in the first relativistic correction, the result (\ref{26}) must be
expanded to $O((Z\al)^4)$.

Unfortunately for hyperfine or Lamb shift energies, the effect of the
primed  states cannot be
removed, since they both come from perturbations 
to the (known) relativistic solution
of the Dirac equation in the presence of a Coulomb potential only. 

A semi-relativistic expression for the Hamiltonian of a 
hydrogenic system was worked
out in Ref. \cite{gabriel}, where the effects of nuclear 
spin (hyperfine effect) were also
included within the context of LLI violations. The result presented there
for the atomic energy levels is incomplete though, since the contribution
of the prime states was overlooked, as discussed at the end of Sec. III.

\setcounter{equation}{0}
\section{Loop calculations\label{Aloop}}

Given the form of the photon propagator (\ref{11}), it is convenient
 to divide the calculation into two parts
\be
\delta E_S= \delta E_S^{(A)}+\delta E_S^{(B)}
\ee
where $\delta E_S^{(A)}$ groups the contributions of the terms proportional to
$\eta_{\mu\nu}$ in  $G_{\mu\nu}$, whereas $\delta E_S^{(B)}$ contains
those proportional to $\gamma^2 = 1/(1-\vec u^2)$ 
and $\xi$. We are interested in solving for the shift in energy levels
up to first order in $\xi$, so it is enough to consider a Coulomb potential
as the source for part B, while for part A the full source as
defined in Eq. (\ref{po}) needs to be included.

We mention again that we are interested in
calculating the GM Lamb shift to lowest nontrivial order in
$\alpha$, {\it i.e.}  up to  $O(\al(Z\al)^4)$.  To this
order, we can use the nonrelativistic expressions for both the
large and small component of the electron spinor
$\psi$. So for example, if we make
the substitution
\be\label{a20} 
\psi(\vec p)=(Z\alpha m)^{-3/2} w(\vec t), 
\ee
where  $w(\vec t)$ is a dimensionless spinor whose first two
components are of order unity, and the last two are of order
$Z\alpha$, we can assign orders to the various terms according to
\ba\label{a20a} 
& &p_i\sim Z\alpha m,\;\;\; E_0-m\sim (Z\alpha)^2m\nonumber\\ 
& &eA_0\,d^3p'\sim eA_i\,d^3p'\sim (Z\alpha)^2m \\ 
& & \sp\,\gamma_i\,\psi_nd^3p\sim Z\alpha m.\nonumber \ea
These approximations will be used in the sequel to simplify the
expressions we obtain.

 \subsection{Type A Contributions to the Self-energy}

Here we will consider
\be\label{a12}
G_{\mu\nu}^{(A)}=-\frac{\eta_{\mu\nu}}{k^2}(1+\xi)
\ee
and $\nn V=-e A_\mu\gamma^{\mu}$, with $ A_\mu$ given by Eq.(\ref{po}).
This part of the calculation is almost identical to that of
BBF\cite{BBF}; 
the only
difference is that now we have to consider a source that 
contains a magnetic part in addition to the electric one.

We begin by computing $\delta E_1$. Relating
the counterterm $\delta C$ to the renormalization of the
electron mass and regularizing the photon propagator via
\be\label{a14}
\frac{1}{k^2}\to-\int_{\mu^2}^{\Lambda^2}\frac{dL}{(k^2-L)^2}.
\ee
we find that $I_2$ and $I_3$ in (\ref{a6}) become
\ba\label{i2i3}
I_2&=(1+\xi)&\nn V\{\half \ln(p^2/\mu^2)-\ln(\Lambda^2/p^2)\}\\
I_3&=(1+\xi)&\frac{3}{4}\{ \nn V (\ln(\Lambda^2/p^2)+\half)+m 
\ln(m^2/p^2)\} \quad . \nonumber 
\ea
On the other hand, we obtain for $I_1$
\ba
I_1&=&(1+\xi)\Big\{-\frac{3}{8}\nn V
-\half p'\cdot p\nn V\int_0^1\frac{dx}{p_x^2}
\ln(p_x^2/\mu^2)+\frac{1}{4}\nn V\int_0^1dx\,\ln(\Lambda^2/p_x^2)\nonumber\\
&+&\half\int_0^1\frac{dx}{p_x^2}\{{(1-x)p^2+xp'^2+2p\cdotp'}\nn V +
\nn p'\nn V\nn p\\
&-&2V\cdot p'(1-x)\nn p-2V\cdot p x\nn p'+V\cdot p_x\nn p_x\}\Big\},\nonumber
\ea
where $p_x=xp'+(1-x)p$. 

We can simplify this expression by letting the
momentum operators $\nn p'$ and $\nn p$ respectively 
act on the spinors $\sp (\vec{p}\,')$ and $\psi(\vec p)$, 
using the Dirac equation and (\ref{a20a}) to keep terms up to
the desired order.

Adding together $I_1$, $I_2$, and $I_3$ 
we obtain a result correct to order $\alpha(Z\alpha)^4$:
\ba\label{a17}
\delta E_1^{(A)}&=&\api (1+\xi)\int\sp(\vec p\,')\left[\nn V\frac{q^2}{m^2}
\left[\frac{1}{3}\ln(\frac{m}{\mu})-\frac{1}{8}\right]
+\frac{i}{4m}q_\nu\sigma^{\mu\nu}V_\mu\right]\psi(\vec p)d^3p'd^3p\nonumber\\
&-&\api (1+\xi)\langle  n|\frac{-3V_0^2+5\vec V^2}{4m}|n\rangle ,
\ea
with $q=p'-p$, and $\sigma^{\mu\nu}=\frac{i}{2}[\gamma^\mu,\gamma^\nu]$.
Note  that the term proportional to $q^2$ in Eq. (\ref{a17}) needs to
be evaluated with only the large component of $\psi$ and
$\nn V\simeq V_0$ ($\gamma_0\sim 1$). 

We point out that the initial ultraviolet divergence in (\ref{i2i3})
is cancelled after the addition of the $I$'s in (\ref{a17}).
The remaining infrared divergence
will be cancelled by a similar term which comes from the many-potential part
of the level shift. A similar cancellation occurs in the non-gauge
invariant term present in Eq. (\ref{a17}). These cancellations are 
non-trivial, and provide useful
cross checks to our calculation.

Consider next the evaluation of $\delta E_2$. Since the
operator $M_\mu$ satisfies the transversality condition
\be
 k\cdot M=k\cdot M^{\dag}=0
\ee
we can write $M_0= \vec k\cdot\vec M/k_0$.

Using
\be
\nn V\nn k\gamma_\mu=2V\cdot k\gamma_\mu-2V_\mu\nn k +\nn k\gamma_\mu\nn V,
\ee
in the first term of Eq.(\ref{a9}) 
the operator $M_\mu^{\dag}$can be decomposed into
\be\label{Msplit}
{M_\mu}^{\dag}={M_\mu}^{\dag I}+{M_\mu}^{\dag II}
\ee
with
\ba\label{mIe}
M_\mu^{\dag I}&=&\{\frac{2p_\mu}{k^2-2p\cdot k}-\frac{2p'_\mu}{k^2-2p'\cdot k}\\
\nonumber&+&\nn k\gamma_\mu(\frac{1}{k^2-2p'\cdot k}-\frac{1}{k^2-2p\cdot k})\}\nn V \\
M_\mu^{\dag II}&=&2(V_\mu\nn k-V\cdot k\gamma_\mu)/(k^2-2p\cdot k), \label{mIIe}
\ea
each of which still satisfies
\be\label{a25}
M^{\dag I}\cdot k=M^{\dag II}\cdot k=0. 
\ee

In terms of these operators we now have
\be\label {a23a}
\delta E_2= \langle  M^IK_+^VM^I\rangle +
\langle  M^{II}K_+^VM^{II}\rangle  + \langle  M^IK_+^VM^{II}\rangle
+\langle  M^{II}K_+^VM^{I}\rangle,
\ee
where  each term represents a contribution to  Eq. (\ref{a8})
involving the products of only $M^I$ or $M^{II}$ or cross terms operators.
The simplification of these terms is quite analogous to that shown in
BBF \cite{BBF}.
The decomposition of the M operator in (\ref{Msplit}) allows one to use
simpler expressions for the bound propagator $K_+^V$. In appendix \ref{Amany} it is
shown that  only in the part $\langle  M^IK_+^VM^I\rangle $ will it be
necessary to use  the bound electron propagator; in all other
contributions it is sufficient to replace $K_+^V$ by the propagator
for free electrons, $K_+^0$. Moreover the main
contribution to $\langle  M^IK_+^VM^I\rangle $ arises from intermediate
states of the electron with nonrelativistic energy so that both $K_+^V$ and
$M^I$ can be replaced by their simpler nonrelativistic approximations. 
It is also shown that the cross term in Eq. (\ref{a23a}) gives a
contribution of  order $\al(Z\al)^5$ and is therefore not relevant in
our calculation.
According to the above considerations we can then approximate Eq. (\ref{a23a}) 
by
\be\label{Ene2}
\delta E_2\simeq\langle  M_{NR}^IK_{NR}^VM_{NR}^I\rangle +\langle  M^{II}K_+^0M^{II}\rangle 
\equiv \langle M^I\rangle+\langle M^{II}\rangle .
\ee

We start evaluating the first term of Eq. (\ref{Ene2}). The nonrelativistic
prescription for $K_+^V$ is given by
\be
K_{NR}^V(x',x)=
\left\{ \matrix{\sum_r\varphi_r(\vec x')\varphi_r^*(\vec x)\exp(-iE_r(t'-t))
&\qquad \hbox{for}\; (t'-t) > 0\cr&\cr 0&\qquad \hbox{for}
\; (t'-t)< 0\cr}\right.
\ee
or in momentum space
\be\label{a28}
K_{NR}^V(E_n-k_0;\vec p\,',\vec p)= -i\sum_r\varphi_r(\vec p\,')\varphi_r^*
(\vec p)(E_r-E_n+k_0)^{-1}
\ee
where $\varphi_r$ represents the large component of the Dirac spinor.

 In the same nonrelativistic approach $M_\mu^I$ reduces to
\be\label{a29}
M_\mu^{I(NR)}\simeq(p_\mu'-p_\mu)\frac{V_0(\vec p\,'-\vec p)}{mk_0}\equiv R_\mu,
\ee
where we have approximated $\nn V\simeq V_0$, because although the magnetic
and electric potential have the same order of magnitude (as powers of
$Z\al$), the $\vec\gamma$ matrix mixes large 
components of
the intermediate states 
with small ones
and therefore introduces corrections one order
higher in $Z\al$. 

Therefore, after replacing Eq. (\ref{a28}) and (\ref{a29}) in Eq. (\ref{a8})
we obtain
\be\label{a29.a}
\langle  M^I\rangle = 
\frac{\al}{4\pi^3 i}\int d^4k\,G^{\mu\nu}(k)
\times \sum_r \frac{\langle  n|R_\mu|r\rangle \langle  r|R_\nu|n\rangle }{k_0-E_n-E_r}
\ee
where we have neglected the contribution of the photon momentum $k$ to
the momentum of the intermediate electron states. This is equivalent
to leaving out the factor $\exp(i\vec k\cdot\vec x)$ in the spatial
integration. This can be done because $k\sim E_n-E_r\sim m(Z\al)^2$,
which is small compared with the electron momentum $\vec p\sim mZ\al$
for nonrelativistic states.

Inserting (\ref{a12}) into (\ref{a29.a}), and using Eq. (\ref{a25}) to 
relate the temporal component of $R$ with its spatial components, which 
satisfy
\be\label{p60}
\langle  n|\vec R |r\rangle =\frac{-1}{m k_0}(E_n-E_r)\langle  n|\vec p|r\rangle ,
\ee
we find, after integration 
\be\label{a30}
\langle  M^I\rangle =\frac{2\alpha}{3\pi m^2}(1+\xi)\sum_r|\langle  n|\vec p|r\rangle |^2\\
     (E_r-E_n)\left[\ln(\frac{\mu}{2|E_n-E_r|})+\frac{5}{6}\right]\nonumber
\ee
where all the states and energies represent  the
non relativistic limit of the Dirac solution.  

Eq.(\ref{a30}) can be simplified by using
\be\label{p62}
\sum_r|\langle n|\vec p|r\rangle |^2(E_r-E_n)=\half\langle  n|\gr^2 V_0|n\rangle ,
\ee
which finally gives
\be\label{a32a}
\langle  M^I\rangle =\frac{\alpha}{3\pi m^2}(1+\xi)\left[\Big (\ln(\frac{\mu}{2E_*})+\frac{5}{6}\Big )
\langle  n|\gr^2V_0|n\rangle +\hat C\right]
 \ee
with 
\be\label{cij}
\hat C\equiv\hat C^{ii},\qquad\hat C^{ij}=2\sum_r\langle r|p_i |n\rangle \langle n|p_j |r\rangle 
(E_r-E_n)\ln|\frac{E_*}{E_n-E_r}|
\ee
where $E_*$ is a reference energy to be defined, and $\hat C^{ij}$ has been introduced
for later convenience.
To obtain this result we have neglected the imaginary part of $\langle 
M^I\rangle $ retaining only the leading terms of  $\langle  M^I\rangle $
in the limit $\mu\to 0$.

In computing  $\langle  M^{II}\rangle $, we can take $K_+^V$ 
to be the free electron propagator, which is
\be
K_+^V(E_n-k_0;\vec p\,'-\vec s\,'-\vec k,\vec p+\vec s-\vec k)=\frac{i\delta^3(
\vec s\,'-\vec p\,'+\vec p+\vec s)}{\nn r-\nn k-m},
\ee
where
\be
r^\mu=(m,\vec s_*),\qquad
\vec s_*=\vec p\,'-\vec s\,'=\vec p+\vec s\nonumber
\ee
upon which $\langle  M^{II}\rangle $ becomes
\be\label{a38}
\langle  M^{II}\rangle =\api\int d^3p'd^3p\,d^3s_*\sp(\vec p\,')V_\alpha(\vec p\,'-\vec s_*)
  N^\alpha_\beta(p_*,s_*)V^\beta(\vec s_*-\vec p)\psi_n(\vec p),
\ee
with
\be\label{a39}
 N^\alpha_\beta(p_*,s_*)=-\frac{4}{i}\int\frac{(\eta^{\alpha\mu}\nn k -k^\alpha\gamma^\mu)
(\nn r-\nn k+m)(\eta_{\beta}^\nu\nn k-k_\beta\gamma^\nu)}
{(k^2-2p_*\cdot k)^2(k^2-2r\cdot k -\vec s_*^2) }G_{\mu\nu}(k)d^4k.  
\ee

In the nonrelativistic domain $\int d^3p V_{\al} \approx (Z\al)^2m$
and so the constant value of $N_{\al}^{\beta}$ (independent
of the momentum and energy of the intermediate states) will already
yield an overall contribution to Eq. (\ref{a38}) of the
desired order $\al(Z\al)^4$.  Note that $N_{\al}^{\beta}$ can be
expanded in powers of the momentum $\vec p\,'$, $\vec p$ or
$\vec s_*$, which are of order $mZ\al$,
and therefore any contribution beyond the constant, $Z\al$-independent
term will be of higher order. The same
argument can be used to neglect the binding energy of the intermediate
states. We can therefore evaluate (\ref{a39}) by
approximating $p\sim p_*$ and $p'\sim p_*$ in the denominator of
$M^{I\dag}$ and $M^I$ respectively, so that $p_*\approx (m,0)$ and 
$s_*\approx 0$. 
 
Evaluating $N$ as in reference \cite{BBF} we find that (\ref{a38}) becomes
\be\label{a40}
\langle  M^{II}\rangle =\api(1+\xi)\langle  n|\frac{-3V_0^2+5\vec V^2}{4m}|n\rangle .
\ee
Note that this term will exactly cancel the non-gauge invariant term
present in  Eq. (\ref{a17}).

Finally we add  Eq.(\ref{a32a}) to Eq.(\ref{a40}) to obtain  $\delta E_2^{(A)}$,
and then add it to Eq. (\ref{a17}) to give the final result for the type-A
contribution to the self-energy:
\ba\label{a42}
\delta E_S^{(A)}&=&\frac{\al}{3\pi m^2}(1+\xi)\Big[ \hat C+\Big(\ln(\frac{m}{2E_*})+
\frac{11}{24}\Big)\langle  n|\gr^2 V_0|n\rangle \nonumber\\
     &+&\frac{3}{4}m\int\sp(\pp\, ')
i\sigma_{\mu\nu} V^\mu q^\nu\psi_n(\pp)d^3p'd^3p\Big] .
\ea

Apart from the constant $(1+\xi)$ factor,
there is no formal difference between the result (\ref{a42}) for this contribution
to the level shift and the standard one \cite{BBF}. However there are
implicit differences which appear in the expression for
$V^\mu$ and the solution for the Dirac 
states $|n\rangle$ (in the non-relativistic
approach here) in the presence of that source.

\subsection{Type B Contributions to the Self-energy}

To solve the type-B contributions we have to consider the photon progator 
\be\label{a64}
G_{\mu\nu}^{(B)}=\xi\frac{\gamma^2}{k^2}\left[\beta_\mu\beta_\nu+\eta_{\mu\nu}
\frac{(\beta\cdot k)^2}{k^2}\right]
\ee
and a source  $A_\mu\simeq\eta_{\mu 0}\phi$.

The evaluation of $\delta E_S^{(B)}$ is achieved by the same procedure as for
part A, where now we use Eq. (\ref{a64}) 
in (\ref{a6}) and (\ref{a8}) to solve for
 $\delta E_1^{(B)}$ and $\delta E_2^{(B)}$ respectively. This
computation is somewhat more laborious than that in part A,
due to the $\beta_\mu\beta_\nu$ tensorial dependence
and the factor $\frac{(\beta\cdot k)^2}{k^2}$
present in this part of the (GM) photon propagator.

To evaluate $I_1$, $I_2$, and $I_3$ we need to modify the BBF technique
by using (\ref{a14}) along with
\be\label{a69}
\frac{1}{k^4}\to-2\int_{\mu^2}^{\Lambda^2}\frac{dL}{(k^2-L)^3}.
\ee
to regulate (\ref{a64}).
The expressions for the $I$'s are somewhat more complicated than those
for $\delta E_S^{(A)}$ (as expected);  but their manipulation and further algebra
follow from BBF \cite{BBF}. The relevant details are in 
appendix \ref{Adetail}; the result for the one potential part is 
\ba\label{a70}
\delta E_1^{(B)}&=&\frac{\al}{3\pi m^2}\gamma^2\xi\int\sp(\vec p\,')\Big\{
\nn V q^2\left[\frac{17}{48}\beta^2-\frac{5}{4}+(\frac{\beta^2}{2}-1)
\ln(\frac{\mu}{m})\right]\nonumber\\
&+&\nn V (\beta\cdot q)^2\left[\frac{5}{6}+\ln(\frac{\mu}{m})\right]\\
&+&(\frac{\beta\cdot p}{2}\nn V-\beta\cdot Vm)i\sigma^{ij}u_i q_j
-m\beta\cdot q  i\sigma^{\mu\nu} V_\mu\beta_\nu
\nonumber\\
&+&m(\frac{\beta^2}{8}-\half)i\sigma^{\mu\nu}V_\mu q_\nu\Big\}
\psi(\vec p)d^3p'd^3p\nonumber\\
&-&\api\gamma^2\xi(1+\frac{7}{8}\beta^2)\langle  n|\frac{V_0^2}{3m}|n\rangle
\nonumber 
\ea
which is good up to order $\al(Z\al)^4$, and we have retained
only the leading terms as $\mu\to 0$. 

The evaluation of $\delta E_2^{(B)}$ is quite analogous to that for
$\delta E_2^{(A)}$. The starting point is Eq. (\ref{Ene2}), 
where $\langle M^I\rangle$
and $\langle M^{II}\rangle$ are still defined 
by (\ref{a29.a}) and (\ref{a38})
respectively. We give calculational details
in appendix \ref{Adetail}, and quote here only the
final result:
\ba\label{a72}
\delta E_2^{(B)}&=&\frac{\al}{3\pi m^2}\gamma^2\xi\Big\{\left[\frac{5}{12}\uu ^2-\frac{1}{12}+
(\half+\frac{\vec u^2}{2})
\ln(\frac{\mu}{2E_*})\right]\langle  n|\gr^2 V_0|n\rangle \nonumber\\
&+&\left[\frac{5}{6}
+\ln(\frac{\mu}{2E_*})\right]\langle  n|(\uu\cdot\vec\gr)^2 V_0|n\rangle
+u_iu_j\hat C^{ij}+(\half+\frac{\vec u^2}{2})\hat C\Big \}\nonumber\\
&+&\api\gamma^2\xi(1+\frac{7}{8}\beta^2)\langle  n|\frac{V_0^2}{3m}|n\rangle 
\ea

We now add (\ref{a70}) to (\ref{a72}) to obtain
\ba\label{a74}
\delta E_S^{(B)}&=&\frac{\al}{3\pi m^2}\xi\Big\{\left[-\frac{11}{12}\uu ^2-\frac{47}{48}+
(\half+\vec u^2)\ln(\frac{m}{2E_*})\right]\langle  n|\gr^2 V_0|n\rangle \nonumber\\
&+&\ln(\frac{m}{2E_*})\langle  n|(\uu\cdot\vec\gr)^2 V_0|n\rangle
+u_iu_j\hat C^{ij}+(\half+\vec u^2)\hat C\\
&+&\int\sp(\vec p\,')\Big[(\frac{\beta\cdot p}{2}\nn V-\beta\cdot Vm)i\sigma^{ij}u_i q_j
+m\vec u\cdot q i\sigma^{\mu\nu} V_\mu\beta_\nu
\nonumber\\
&-&m(\frac{\vec u^2}{2}+\frac{3}{8})i\sigma^{\mu\nu}V_\mu q_\nu\Big]
\psi(\vec p)d^3p'd^3p\nonumber\Big\}
\ea
where we approximated  $\gamma^2\simeq1+\vec u^2$ in order to keep
terms only up to order $\vec u^2$. As a cross-check on the above
result we note that, before expanding $\gamma^2$, 
the limit $\beta_\mu\beta_\nu\rightarrow
\eta_{\mu\nu}$, yields
$\delta E_S^{(B)}\rightarrow -2\xi\gamma^2\delta E_S^0$. This is as
expected since according to (\ref{a64}), $G_{\mu\nu}^{(B)}
\rightarrow-2\xi\gamma^2G_{\mu\nu}^0$, where $G_{\mu\nu}^0$ is
the standard (metric) propagator.

We close this section with a comment on the renormalization 
procedure. For $\delta E_S^{(A)}$, the counterterm $\delta C$ was
related to mass renormalization. However in this part of the
calculation we must also account for the renormalization
of the \tmu parameters, which show up as functions of: the limiting speed for
massive particles, $c_0^2\equiv T_0/H_0$, and the photon velocity, $c_*^2\equiv1/\mu_0
\epsilon_0$. Charge renormalization is not necessary here
because the Ward Identity forces a cancellation
between the divergences coming from the one potential part and many potential part of
the self energy. 
Details of this process are shown in appendix \ref{Arenor}.

\subsection{Vacuum Polarization}

We now need to obtain the vacuum polarization contribution. To the desired
approximation, the electrons forming the loop in diagram 1(b) can be considered
free. This is because Furry's theorem implies that
the next-order correction to this is a
diagram which contains a loop with 4 vertices, which is expected to be of order
$\al(Z\al)^6$.
 In that case the result is known to be
\be\label{ap}
\delta E_P=\int\sp(\vec p\,')i\Pi^{\mu\nu}(q)iG_{\nu\sigma}(q)\gamma^\sigma
V_\mu(\vec q)\psi(\vec p)d^3p'd^3p,
\ee
The evaluation of $\Pi^{\mu\nu}$ is identical to the standard (metric)
case, since it only involves the product of fermion propagators, which are
unchanged by the \tmu action. The differences appear in the
renormalization process, where both the charge and the \tmu parameters
must be renormalized, the details of which are shown in appendix
\ref{Arenor}. The result is
\be\label{pol}
\Pi^{\mu\nu}(q)\simeq-\frac{\al}{15\pi}\frac{q^2}{m^2}(q^2\eta^{\mu\nu}-q^\mu q^\nu)
\ee
If we substitute Eqs. (\ref{11}) and (\ref{pol}) in (\ref{ap}), we obtain after some
manipulation 
\be\label{pola}
\delta E_P=\frac{\al}{3\pi m^2}\Big\{\langle n|\gr^2 V_0|n\rangle\Big(-\frac{1}{5}
+\xi\frac{\vec u^2}{5}\Big)-\frac{\xi}{5}\langle n|(\vec u\cdot\vec\gr)^2 V_0)|n\rangle\Big\}
\ee

We next proceed to add together the self energy and vacuum
polarization contributions to the level shift.

\subsection{The total GM Radiative Correction}

 Up to this point we have been able to solve the level shift in terms of
\be\label{abp}
\delta E_n=\delta E_S^{(A)}+\delta E_S^{(B)}+\delta E_P
\ee
where each term has been defined in Eqs. (\ref{a42}), (\ref{a74}) and (\ref{pola}).

We  note that in $\delta E_S$ there are terms proportional to $\vec\gamma$,
which mix large ($\varphi$) and small component 
($\chi$) of $\psi$. Within the accuracy required
we can relate them by $\chi=-i\frac{\vec\sigma\cdot\vec \gr}{2m}\varphi$, and so
write everything in terms of the large component only.

Replacing the expression for the external source (\ref{po}) 
in  (\ref{abp}), we obtain after some algebra
\be\label{Elamb}
\delta E_n=\frac{\alpha}{3\pi m^2}\left[ \Big (1+\xi (\frac{3}{2}+\vec u^2)\Big )\hat C
+\xi u_iu_j \hat C^{ij}+\langle n|\hat E|n\rangle\right]
\ee
where $\hat C$ and $\hat C^{ij}$ are defined by Eq. (\ref{cij}), and
\ba\label{hate}
\hat E&=&4\pi Z\alpha\delta (\vec x)\left[\frac{19}{30}+\ln (\frac{m}{2E_*})+
\xi\Big[-\frac{1}{30}-\frac{58}{45}\vec u^2+(\frac{3}{2}+\frac{2}{3}\vec u^2)
\ln(\frac{m}{2E_*})\Big]\right]\nonumber\\
&+&3\frac{Z\alpha}{r^3}\left[\frac{1}{4}+\xi\Big[\frac{1}{8}
-\frac{\vec u^2}{2}-(\vec u\cdot\nv)^2\Big]\right]\vec \sigma\cdot\vec L\\\nonumber
&-&\xi\frac{Z\alpha}{r^3} [3(\vec u\cdot\nv)^2-\vec u^2]\left[\frac{14}{15}
+2\ln (\frac{m}{2E_*})\right]\\\nonumber
&+&\frac{\xi}{2}\frac{Z\alpha}{r^2}\left[
\frac{7}{2}\vec u\cdot\nv\,\vec\sigma\cdot(\vec u\times\vec p)
-\vec\sigma\cdot(\vec u\times\nv)
\vec u\cdot\vec p\right]
\ea
We have omitted operators with odd parity (such as 
$\vec u\times\nv\cdot\vec\sigma$) in (\ref{hate}), 
since their  expectation values vanish for states of definite parity. 

There is still an implicit dependence on $\xi$ and $\vec u$ in
(\ref{Elamb}), which
comes from the Dirac states  (as seen at the end of Sec. III). Note that up
to this order all  atomic states and energies refered in Eqs. (\ref{Elamb}) and
 (\ref{cij}) are considered within a non relativistic approach.

In terms of the formal solution for the Dirac equation (\ref{solution}), we
can single out the complete $\xi$ dependence in  (\ref{Elamb}), and write
\be\label{efinal}
\delta E_n=\frac{\alpha}{3\pi m^2}\Big\{ \Big (1+\xi (\frac{3}{2}+\vec u^2)\Big )\hat C^0
+\xi u_iu_j \hat E^{ij}+\,^0\langle n|\hat E|n\rangle ^0\Big\}
\ee
with 
\be\label{Eij}
u_iu_j \hat E^{ij}=u_iu_j \hat C^{ij}+\hat C'+(\,^0\langle n|\hat E_{\xi=0}|n\rangle'+\hbox{h.c.})
\ee
where $\hat C'$ groups all the terms in Eq. (\ref{cij}) depending on  the perturbative
states ($|n\rangle'$) or energies ($E_n'$) as introduced in
Eq. (\ref{solution}). These perturbative states  
are needed not only for the $|n\rangle$ state
related to the level shift, but for all the intermediate states
introduced by (\ref{cij})as well. 
Eq. (\ref{efinal}) is valid up to $O(\xi)O(\vec u^2)O(\al(Z\al)^4 )$.

We can define the reference energy $E_*$ as in the metric case by \cite{IZ}
\be\label{erefer}
\left.\begin{array}{ll}
\ln(E_*^{n0})
=\frac{\sum_r|\langle  r|\vec p|n\rangle |^2(E_r-E_n)\ln|E_r-E_n|}{\sum_r|
\langle  r|\vec p|n\rangle |^2(E_r-E_n)}
&\mbox{for $l=0$}\\&\\
2\frac{m^3}{n^3}(Z\al)^4\ln(\frac{Z^2 Ryd}{E_*^{nl}})=\sum_r|\langle  r|\vec p|n\rangle |^2
(E_r-E_n)\ln|\frac{1}{E_r-E_n}|
&\mbox{for $l\not=0$}
\end{array}\right.
\ee
where the subscript $0$ has been omitted in the energies and states. This
definition reduces
\be\label{C0}\hat C^0=\left\{\begin{array}{ll}
0&\hbox{for $l=0$}\\
4\frac{m^3}{n^3}(Z\al)^4\ln(\frac{Z^2 Ryd}{E_*^{nl}})&\hbox{for $l\not=0$}
\end{array}\right.
\ee
which provides an elegant way to write the ``Bethe-sum".
The presence of prefered
frame effects will induce more ``Bethe-sum"--like terms in
$\hat C^{ij}$ which, along with the contribution 
from the perturbative states (both
ones counted by $\delta \hat E^{ij}$) will have to be evaluated
numerically for any particular state. 

 For the Lamb shift states we can use \cite{IZ} :
\be\label{refener}
 E_*^{2S}=16.640\,\hbox{Ryd}\qquad E_*^{2P}=0.9704\,\hbox{Ryd} 
\ee
and simplify the last term in Eq. (\ref{efinal}) as
\ba\label{E0}
^0\langle\hat E\rangle ^0_{2S_{1/2}}\!\!&=&\!\!\frac{(Z\al)^4}{2}m^3\left\{
\frac{19}{30}+\ln (\frac{m}{2E_*^{2S}})-
\xi\Big[\frac{1}{3}+\frac{58}{45}\vec u^2-(\frac{3}{2}+\frac{2}{3}\vec u^2)
\ln(\frac{m}{2E_*^{2S}})\Big]\right\}\\\nonumber
^0\langle\hat E\rangle ^0_{2P_{1/2}}\!\!&=&\!\!\frac{(Z\al)^4}{2}m^3\left\{
-\frac{3}{24}-\frac{\xi}{12}\left[\frac{3}{4}
-\vec u^2\left[\frac{107}{30}-\frac{1}{6\sqrt{10}}
+\cos^2\theta\left(\frac{1}{12}+\frac{1}{6\sqrt{10}}\right)\right]\right]\right\}
\ea
where $\theta$ represents the angle between the atom's quantization axis and the
frame velocity $\vec u$.

\subsection{Many potential part approximations\label{Amany}}

In this appendix we justify the following approximations:
\ba\label{ap1}
\langle M^IK_+^VM^I\rangle&\simeq&\langle M_{NR}^IK_{NR}^VM_{NR}^I\rangle\\\label{ap2}
\langle M^{II}K_+^VM^{II}\rangle&\simeq&\langle M^{II}K_+^0M^{II}\rangle\\\label{ap3}
\langle M^I K_+^V M^{II}\rangle&\simeq&O((Z\al)^5\al)
\ea
following arguments similar to those presented by BBF \cite{BBF}.

We first note that, as powers of $Z\al$, the orders of magnitude
of the different terms involved in the expressions in (\ref{ap1}) are equivalent
to those for the metric case.  For example, if we look at the source, we see that 
$eA_\mu\sim e\phi $,
where $A_\mu$ is given by Eq. (\ref{po}) and $\phi$ is the ordinary Coulomb potential, 
and so the
relative order between the nonmetric and metric case is the same. Furthermore,
as discussed at the end of Sec. III, the states $|n\rangle$ and $|n\rangle ^0$ also have the same 
order of magnitude,  as do the quantities $E_n$ and $E_n^0$.  Discrepancies that
could be expected from the photon propagator, particularly from the part proportional to
$\beta^\mu\beta^\nu$ (in contrast to the $\eta_{\mu\nu}$ dependence for the
standard case), are not important as long as the transversality condition
is satisfied for the $M$ operators, since this condition relates the differing components with
the appropriate orders of magnitude. Finally, unlike the photon propagator, the bound
propagator retains the same form as in the standard case, with differences arising only from the expression
for the external source. As a consequence its further simplification is analogous to the 
metric (BBF) case.

Let us look at the many potential part. From (\ref{a8}) we get
\ba\label{a88}
\langle  M K_+^V M \rangle
&=&\int \sp_n(\vec p\,')M_\mu(p',p'-s'-k)\\
 &\times&K_+^V(E_n-k_0;\vec p\,'-\vec s\,'-\vec k,\vec p+\vec s-\vec k)\nonumber
\\
 &\times&M_\nu^{\dag}(p+s-k,p)\psi_n(\vec p)G^{\mu\nu}(k)\nonumber
\ea
for the generic structure of the terms on the 
left hand sides of (\ref{ap1})--(\ref{ap3}),
where the constant factors and 
integrations over $p_i$ and $s_i$ have been omitted.
The nonrelativistic and relativistic regions are defined
according to $|\vec{k}|\sim(Z\al)^2m<<m$ and $|\vec{k}|>m$,respectively.
In considering the relevant orders of magnitude in each of the expressions 
(\ref{ap1})--(\ref{ap3}) that follow from (\ref{a88}), we note that,
to lowest order in $Z\al$, 
the relevant contribution from $G^{\mu\nu}$ comes when $k_0\sim |\vec{k}|$, 
and that  we can employ the nonrelativistic expressions
for the $\psi_n$, making use of the approximations given by (\ref{a20a}).

Turning now to the relation (\ref{ap1}),  we can prove it 
by showing that the contribution of
relativistic states for $M^I$ is of a 
higher order of magnitude than for $M^{II}$.  
We can see from (\ref{mIe}) and (\ref{mIIe}) that 
 $M^I$ differs from $M^{II}$ by a factor (leaving aside the temporal component)
$(\vec{p}^\prime - \vec{p})/k_0$, 
which in the relativistic region ($k_0 \sim m$) is of
order $Z\al$. Therefore the contribution of $M^I$ in that domain will be of
at least one order higher than that of $M^{II}$. Since
the latter is already of the desired order 
(assuming the validity of (\ref{ap2}) )
we can neglect the contribution
of the relativistic states for $M^I$, and consider it, along with  the
bound propagator, in its nonrelativistic limit.

To prove the relation (\ref{ap2}) we evaluate the 
error due to the neglect of the 
electromagnetic potential in the 
intermediate states. We imagine that one extra 
potential ($\nn V$) acts between $M^{II\dag}$ and $M^{II}$. This introduces an
extra factor of order
\be
\int d^3r'\frac{\nn V(r'-r)}{\nn r'-\nn k-m}\sim\int d^3r'\frac{\nn V(r'-r)}{2k_0 m}\nn k
\sim (Z\al)^2
\ee
which is negligible within the accuracy required. We have then shown that, in the evaluation
of $M^{II}$, the intermediate states may indeed be regarded as free.

The relation (\ref{ap3}) follows from arguments 
similar to those used to justify 
(\ref{ap1}). Since in the relativistic region $M^{I}$ is one order
higher than $M^{II}$, the cross term in that region will also be
one order higher than $\langle M^{II}\rangle$, and so is negligible. 
On the other hand
in the nonrelativistic region $M^{I}$ will be dominant 
(note the factor $k_0$ in its denominator) over $M^{II}$. That is
\be
|\frac{M^{II}}{M^{I}}|\sim |\frac{k_0}{\vec p'-\vec p}|\sim Z\al
\ee
and so the product of these terms will be negligible in comparison with 
$\langle M^{I}\rangle$. Hence the cross terms yield results that are at least
one order higher than the desired order, 
and so they do not need to be included.

\subsection{Renormalization\label{Arenor}}

Just as in the standard (metric) case, we need to renormalize the
various parameters of the theory in order to get rid of the
divergences. In the standard case, those parameters are the
mass and charge, although the latter only needs to be renormalized
for the vacuum polarization contribution. The self energy part has no
 need for such a renormalization, since the divergences coming from the one potential
and many potential parts cancel each other. In the nonmetric
case, we have also to include the renormalization of the \tmu parameters,
which show up as functions of $c_0^2\equiv T_0/H_0$
and $c_*^2\equiv 1/\mu_0\epsilon_0$. 

In part A of the calculation, renormalization is identical
to the standard case. The counterterm $\delta C$ is just related to mass renormalization.
In part B, we need to consider additional counterms, since
$\delta C$ should also account for the renormalization of the
\tmu parameters.

In units where $c_0\equiv 1$ ($c_*=c$), EEP-violating corrections only appear in the electromagnetic sector of 
the action (as terms proportional to $\xi$). However we could choose more generally
$c_0\not =1$, for which the particle sector of the Lagrangian density is of the form 
\be
{\cal L}_D=\sp(\nn p-\nn V-m)\psi+\xi_0\sp(p_0-A_0)\gamma^0\psi
\ee
with $\xi_0\equiv 1-c_0^{-1}$; or in the moving frame (after using (\ref{8})) is
\ba \label{apbD}
{\cal L}'_D&=&\sp(\nn p-\nn V-m)\psi\\
&+&\xi_0\gamma^2\sp(\beta\cdot p-\beta\cdot V)\nn \beta\psi\nonumber
\ea
up  to a constant.

 From (\ref{apbD}) we see that quantum corrections of the form
\be
\delta {\cal L}_D=\sp(\delta\xi_0^{(1)}\beta\cdot p-\delta\xi_0^{(2)}
\beta\cdot V)\nn\beta\psi
\ee
can still be expected. Note that gauge invariance will guarantee $\delta\xi_0^{(1)}=
\delta\xi_0^{(2)}=\delta\xi_0$.
Hence, in order to renormalize the mass and the \tmu parameters, we have to
include counterterms of the form
\be
\delta C=\delta m+\delta\xi_0\nn\beta(\beta\cdot p-\beta\cdot V)
\ee
where $\delta m$ and $\delta\xi_0$ are chosen such that $\delta E_S$
gives zero contribution as the source is turned off. This condition forces
$I_3=0$ when acting on free spinors.

Finally, for  the vacuum polarization contribution the charge
has to be renormalized along with the \tmu parameters. Charge
renormalization is identical to the standard case. For the 
\tmu parameters the procedure is equivalent to the self energy part,
where now, given the form of the electromagnetic
action (see Eq. (\ref{9})), we expect quantum fluctuations of the form
\be
\delta {\cal L}_{EM}=\delta \xi A^\mu\{(k^2-(\beta\cdot k)^2)\eta_{\mu\nu}
-\beta_\mu\beta_\nu k^2\}A^\nu
\ee
to occur. Hence a 
counter term of that form it is needed to renormalize the
\tmu parameters, or equivalently $\xi\equiv 1-H_0/T_0\mu_0\epsilon_0$.

\subsection{Calculational Details of Type B Contributions\label{Adetail}}

We present here further details underlying
the computation leading to Eqs. (\ref{a70}) and (\ref{a72}),
which are referred as the type-B contributions to the self energy. 
In this part the photon
propagator to be considered is given by (\ref{a64}), where the
first and second terms have
respectively a tensor dependence like 
$\beta_\mu\beta_\nu$ and $\eta_{\mu\nu}$, and need
to be regularized according to (\ref{a14}) 
and (\ref{a69}). We show the relevant details
involving the first term of the propagator only, since the remainder 
can be computed in a similar way.
 
We begin then with the one potential part by simplifying $I_1$. 
After replacing (\ref{a64}) in (\ref{a7}), we get
\be\label{c1}
I_1=-\frac{i}{4\pi^2}\gamma^2\xi\int\frac{2p'\cdot\beta-\nn\beta\nn k}{k^2-2p'\cdot k}
\nn V\frac{2p\cdot\beta-\nn k\nn\beta}{k^2-2p\cdot k}\frac{d^4k}{(k^2-L)^2}dL+\cdots
\ee
where from now on $\cdots$ stands for the contributions coming from the second term of (\ref{a64}).

If we use
\be\nonumber
\frac{1}{abc^2}=6\int_0^1dx\int_0^1\frac{z(1-z)dz}{[(ax+b(1-x))(1-z)+cz]}
\ee
we can rewrite Eq. (\ref{c1}) as
\ba\nonumber
I_1=&-&4p\cdot\beta p'\cdot\beta\nn V J_0+2p\cdot\beta\nn\beta\gamma^\mu\nn V J_\mu\\
&+&2p'\cdot\beta\nn V\gamma^\mu\nn\beta J_\mu-\nn\beta\gamma^\mu\nn V\gamma^\nu\nn\beta J_{\mu\nu}
+\cdots\nonumber
\ea
where
\be\label{c3}
J_{\{0;\mu;\mu\nu\}}=-\frac{3i}{2\pi^2}\gamma^2\xi\int_0^1dx\int_0^1z(1-z)dz\frac{dLd^4k}{[(k-p_x(1-z))^2-
\Delta_L]^4}\{1;k_\mu;k_\mu k_\nu\}
\ee 
with
\be
p_x=xp'+(1-x)p\qquad \Delta_L=p_x^2(1-z)^2+Lz
\ee
After evaluating (\ref{c3}), we can express
\ba
I_1&=&\gamma^2\xi\int \frac{dx}{p_x^2}\Big\{\nn V\Big[\half\beta\cdot p\beta\cdot p'
(\ln\frac{p_x^2}{\mu^2}-2)+\frac{\beta^2}{8}p_x^2(\frac{3}{2}-\ln\frac{\Lambda^2}{p_x^2})\Big]\nonumber\\
&-&x\frac{\nn p'}{2}\Big(\beta\cdot p\nn V\nn\beta+\beta\cdot p'\nn\beta\nn V\Big)
-\Big(\beta\cdot p\nn V\nn\beta+\beta\cdot p'\nn\beta\nn V\Big)(1-x)\frac{\nn p}{2}\nonumber\\
&+&\nn p_x\Big(\beta\cdot V(\beta\cdot p +\beta\cdot p')-\frac{\beta^2}{4}p_x\cdot V\Big)\\
&+&\nn\beta\Big(\half p_x\cdot\beta p_x\cdot V-\frac{1}{4} V\cdot\beta p_x^2(1+\half
 -\ln\frac{\Lambda^2}{p_x^2}\nonumber\\
&-&(1-x)\beta\cdot p V\cdot p-x p'\cdot\beta V\cdot p'\Big)\Big\}+\cdots\nonumber
\ea

The evaluation of the remaining $I$'s is analogous, and so
\ba
I_2&=&\nn V\gamma^2\xi\Big\{\frac{\beta^2}{4}(\ln\frac{\Lambda^2}{p^2}-1)+
\frac{(\beta\cdot p)^2}{p^2}(1+\half \ln\frac{\mu^2}{p^2})\Big\}+\cdots\\
I_3&=&-\frac{1}{4}\beta\cdot p\nn\beta\gamma^2\xi(\ln\frac{\Lambda^2}{p^2}+
\frac{5}{2})-\frac{1}{8}\nn p\beta^2\gamma^2\xi(\ln\frac{\Lambda^2}{p^2}-\half)+\cdots+
\delta C
\ea

{}From appendix \ref{Arenor}, we know 
$\delta  C=\delta m+\delta\xi_0\nn\beta(\beta\cdot p-\beta\cdot V)$,
where in this case
\be\nonumber
\delta m=\frac{\beta^2}{8}\gamma^2\xi\Big(\ln\frac{\Lambda^2}{m^2}-\half\Big)+\cdots\qquad
\delta\xi_0=\frac{1}{4}\gamma^2\xi\Big(\ln\frac{\Lambda^2}{m^2}+\frac{5}{2}\Big)+\cdots
\ee

Since here $V^\mu=\eta^{\mu 0} V_0$, we can rewrite after some manipulation
\be
I_1+I_2+I_3=\gamma^2\xi(K_1+K_2+K_3)+\cdots
\ee
where
\ba
K_1&=&\nn V\int\frac{dx}{p_x^2}\left[\Big(\ln(\frac{\mu}{E_0})+1\Big)\left(\frac{(\beta\cdot p)^2}{p^2}
-\beta\cdot p\beta\cdot p'\right)+\half\beta\cdot p\beta\cdot p'\ln(\frac{p_x^2}{E_0^2})\right]\nonumber\\
K_2&=&\Big(m\frac{\beta^2}{2}+\beta\cdot p\nn\beta\Big)\left[-\frac{1}{4}\ln(\frac{m^2}{p^2})
\delta(p-p')-\frac{\nn V}{2m}-\frac{1}{4}\beta\cdot V\nn\beta \ln(\frac{p^2}{m^2})\right]
\nonumber\\\nonumber
K_3&=&-\half\int \frac{dx}{p_x^2}\Big\{ x\beta\cdot p'\nn p'\nn\beta\nn V+(1-x)\beta\cdot p\nn V\nn\beta
\nn p\\\nonumber
&+&V\cdot p\left(\frac{\beta^2}{2}\nn p_x+\beta\cdot p_x\nn\beta\right)
-2\beta\cdot V(\beta\cdot p+\beta\cdot p')\nn p_x+x\beta\cdot p\nn p'\nn V\nn\beta\\\nonumber
&+&(1-x)\beta\cdot p'\nn\beta\nn V\nn p-p_x^2\left[\frac{\beta^2}{2}\nn V-2\beta\cdot V\nn\beta+\nn V\nn\beta
\frac{\beta\cdot p}{m}\right]
\Big\}
\ea

We want a result good to $\al(Z\al)^4$, 
and so we can simplify the above expressions by
using the assigned order given by (\ref{a20a}), from which we can relate
\be\nonumber\begin{array}{l}
q\equiv p'-p\sim Z\al m\\
p'^2-p^2\sim(\beta\cdot p)^2-p^2\sim p_x^2-m^2\sim (Z\al)^2 m^2
\end{array}
\ee
and then reduce $K_1$ to
\be
K_1\simeq\frac{\nn V}{2m^2}\left\{\left[(\beta\cdot q)^2-\frac{q^2}{3}\right]\ln(\frac{\mu}{m})
-\frac{5}{6}q^2(\half+\beta^2)+(\beta\cdot q)^2\right\}
\ee
where antisymmetric terms under $p'\leftrightarrow p$ vanish.

To simplify $K_2$ we follow BBF and use
\ba
(\nn p-m)^2\nn\beta&=&\nn V(\nn p-m)\nn\beta=2\beta\cdot p\nn V-2m\nn V
\nn\beta-\nn V\nn\beta\nn V\nonumber\\
&\simeq&2(V\cdot p-m\nn V)\nn\beta-\nn q\nn\beta\nn V- V^2
\ea
where we have assumed the operator is acting on Dirac spinors of momentum $p$
and omitted the integration coming from
\be\sp(p)(\nn p-m)=\delta(q_0)\int\sp(p')\nn V(q)d^3p'\ee

Note that $\nn V\nn\beta\nn V\simeq V^2$, since the square of the potential 
(after factoring out the spinors and integration variables) is already of the
desired order $(Z\al)^4$ (see (\ref{a20a})) and so $\nn\beta\simeq\gamma_0
\simeq 1$.

The final result is
\be
K_2\simeq-\frac{V^2}{4m}(\frac{\beta^2}{2}+5)+(V\cdot p-m\nn V)\frac{\nn\beta}{2m}
-\frac{\beta\cdot p}{4m^2}\nn q\nn\beta\nn V
\ee

Following a similar approach we reduce
\ba
K_3&\simeq&-\frac{\nn\beta}{2m}(V\cdot p- m\nn V)+\frac{V^2}{m}+\beta\cdot V \frac{\beta\cdot q}
{m}-\frac{\beta\cdot q}{4m^2}\nn\beta V\cdot p\\
&-&\frac{\beta\cdot q}{2m}\nn\beta\nn V-\frac{\beta\cdot V}{2m}\nn q\nn\beta
+\frac{\beta^2}{8m}\nn q\nn V-\frac{q^2}{12m^2}(\frac{\beta^2}{2}-1)\nn V\nonumber
\ea

We can  make further simplifications by using
\be
\int\sp(p')B(p',p)\psi(p)d^3p'd^3p=0,
\ee
provided $\gamma^0 B^{\dag}(p',p)\gamma^0=-B(p,p')$, 
where $B$ represents any operator
as a function of $p'$ and $p$, as for example, $\beta\cdot q\nn V$. Note that we are interested
only in the real part of the level shift. 

Putting everything together, we obtain after some manipulation

 \ba\label{aE1}
\delta E_1^{(B)}&=&\frac{\al}{\pi m^2}\gamma^2\xi\int\sp(\vec p\,')\Big\{
\nn V (\beta\cdot q)^2\left[\frac{5}{8}+\half\ln(\frac{\mu}{m})\right]
-\nn V q^2\left[\frac{\beta^2}{16}+\frac{1}{8}+\frac{1}{6}
\ln(\frac{\mu}{m})\right]\nonumber\\
&-&(\frac{\beta\cdot p}{4}\nn V+\beta\cdot V\frac{m}{2})i\sigma^{ij}u_i q_j
-\frac{m}{2}\beta\cdot q  i\sigma^{\mu\nu} V_\mu\beta_\nu\\
&+&m(\frac{\beta^2}{8}-\half)i\sigma^{\mu\nu}V_\mu q_\nu\Big\}
\psi(\vec p)d^3p'd^3p
-\api\gamma^2\xi(1+\frac{\beta^2}{2})\langle  n|\frac{V_0^2}{4m}|n\rangle+\cdots
\nonumber 
\ea

Note again that this represents the calculation involving only the first term of  Eq. (\ref{a64}). 

Now to evaluate the many potential part contribution we need to solve Eq. (\ref{Ene2}), with
$\langle M^I\rangle$ and $\langle M^{II}\rangle$ given by Eqs. (\ref{a29.a}) and (\ref{a38})
respectively.

So, after substituting (\ref{a64}) in (\ref{a29.a})
\be\label{dd1}
\langle M^I\rangle=\frac{\al}{4\pi^3i}\gamma^2\xi\sum_r\int\frac{|\beta\cdot{\cal M}|^2}{k^2-\mu^2}
\frac{d^4k}{k_0-E_n-E_r}+\cdots
\ee
with
$${\cal M}_\mu\equiv\langle n |R_\mu|r\rangle,$$

Using the transversality condition, we relate
$$ {\cal M}_0=\frac{\vec k\cdot\vec{\cal M}}{k_0}=\frac{|\vec k|}{k_0}|\vec{\cal M}|
\cos\theta$$
which reduces the integral on the angles of $\vec k$ to
\be
\int d\Omega|\beta\cdot{\cal M}|^2=4\pi\left(\frac{\vec k^2}{3k_0^2}|\vec{\cal M}|^2
+|\vec u\cdot\vec{\cal M}|^2\right)
\ee

We  evaluate the remaining $k_0$ and $|\vec k|$ integrations in (\ref{dd1}), by
using (\ref{p60}), (\ref{p62}) along with the analogous relations
\ba
\vec u\cdot\vec{\cal M}&=&\frac{-1}{mk_0}(E_r-E_n)\langle n |\vec u\cdot\vec p|r\rangle
\nonumber\\
\sum_r|\langle n|\vec u\cdot\vec p|r\rangle|^2(E_r-E_n)&=&\half\langle n|\vec u\cdot\vec\gr V_0|n\rangle
\nonumber
\ea
to finally obtain
\ba\label{mm1}
\langle M^I\rangle=\frac{\al}{\pi m}\gamma^2\xi&\Big\{&\frac{1}{6}\hat C+\half u_i u_j
\hat C^{ij}+\left[\frac{2}{9}+\frac{1}{6}\ln(\frac{\mu}{2E_*})\right]\langle  n|\gr ^2V_0|n\rangle\\
\nonumber&+&
\left[\frac{1}{2}+\frac{1}{2}\ln(\frac{\mu}{2E_*})\right]\langle  n|(\vec u\cdot\vec\gr) ^2V_0|n\rangle
\Big\}+\cdots
\ea
where we have kept only the leading terms as $\mu\rightarrow 0$ and neglected the imaginary part.

The computation of $\langle M^{II}\rangle$ is straightforward. 
Here we need to replace (\ref{a64}) in (\ref{a38}),
and use $V_\alpha=\eta_{\al 0} V^0$. Further 
simplifications follow from BBF and the assigned
order of magnitude given before. The final result is
\be\label{mm2}
\langle M^{II}\rangle=\frac{\al}{\pi}\gamma^2\xi(\frac{\beta^2}{2}+1)\langle n |\frac{V_0^2}{4m} |n\rangle+\cdots
\ee

Adding together (\ref{aE1}), (\ref{mm1}), and (\ref{mm2}) will give us then the final expression
for the self energy contribution for this part of the calculation. Note that the above results
can be verified by taking the limit $\beta_\mu\beta_\nu\rightarrow\eta_{\mu\nu}$, which reduces
$$G_{\mu\nu}^{(B)}\rightarrow-\gamma^2\xi G_{\mu\nu}^0+\cdots,$$ and therefore
the former expressions should reduce  up to a constant, to the metric case.

\setcounter{equation}{0}
\section{Virtual non metric anomaly\label{Apol}}
In the \tmu formalism, gravity interacts with matter through the $T$ and $H$ functions, which
are assumed locally constant within atomic scales. A priori they do not need to be the
same for differents type of matter (like baryons and leptons), or furthermore for
matter and antimatter. In this context for example, a non metric anomaly
related to electron/ positron difference will modified the Lagrangian density related 
to fermions by
\be\label{posi1}
{\cal L}_D=\sp(\nn p-\nn V-m)\psi+\xi_+\sp^+(p_0-A_0)\gamma^0\psi^+
\ee
where $\xi_+\equiv 1-c_-/c_+$ and $c_\mp=(T_\mp/H_\mp)^{1/2}$, with $-$ and
$+$ labeling electrons and positrons respectively. 
After using (\ref{8}), we can refer (\ref{posi1}) to the moving frame as
\be \label{posi2}
{\cal L}'_D=\sp(\nn p-\nn V-m)\psi
+\xi_+\gamma^2\sp^+(\beta\cdot p-\beta\cdot V)\nn \beta\psi^+
\ee

The imposed broken symmetry between particle and antiparticle changes the
fermion  propagator (in the positron case) to (up to $O(\xi_+)$):
\be\label{posi3}
S_F^{+}=(\nn p-m)^{-1}+\xi_+(\nn p-m)^{-1}\gamma^2\nn\beta\beta\cdot p(\nn p-m)^{-1}
\ee

where the first term represents the unchanged electron propagator $S_F^-$.

The positron-electron pairs  produced in the electric field of the atomic
nucleus,  are seen in the Lamb shift transition via the vacuum polarization
contribution given by (\ref{ap}), where in this case:
\be\label{posi4}
i\Pi^{\mu\nu}(q)=\frac{(ie)^2}{(2\pi)^4}(-1) Tr\int d^4p\gamma^\mu  i S_F^-(p+q)
\gamma^\nu i S_F^+(p)
\ee

After using eq. (\ref{posi3}) along with standar technics \cite{IZ}, we obtain that
the non metric part of (\ref{posi4}) is up to $O(q^2)$
\be\label{posi5}
i\Pi^{\mu\nu}(q)^+=-\frac{\al}{2\pi}\gamma^2\eta^{\mu\nu}\frac{q^2}{m^2}\left\{\frac{1}{30}
q^2\beta^2-\frac{1}{5}(\beta\cdot q)^2\right\}+\cdots
\ee
where $\cdots$ accounts for the gauge dependent terms which give no
contribution to (\ref{ap}).  Eq (\ref{posi5}) also comes after proper
regularization and renormalization processes, which follow from
previous sections.

In this EEP violating context, the radiative corrections related to atomic
energy levels are modified by (up to $O(\al(Z\al)^4\,O(u^2)$)
\be
\delta E_L^+=\delta E_P^+=-\xi_+\frac{\al}{10\pi m^2}\Big\{\frac{1}{6}\langle n|\gr^2 V_0|n\rangle
+\langle n|(\vec u\cdot\vec\gr)^2 V_0)|n\rangle\Big\}
\ee
where we have replaced (\ref{posi5}) in (\ref{ap}) and simplified afterwards.
 By taking the Lamb atomic states, we finally obtain
\be
\Delta E_L^+=-\xi_+\frac{m}{120\pi} (Z\al)^4\al  (1+2\vec u^2)
\ee

\end{document}